\documentstyle[prl,twocolumn,aps,epsf,eqsecnum,bbox]{revtex}
\begin{document}


\voffset1.5cm


\title{Transverse dynamics of hard partons in nuclear media\\
and the QCD dipole}
\author{Urs Achim Wiedemann}

\address{Theory Division, CERN, CH-1211 Geneva 23, Switzerland}

\date{\today}
\maketitle

\begin{abstract}
We derive the non-abelian generalization of the Furry approximation
which describes the transverse dynamical evolution of a hard 
projectile parton inside a spatially extended 
colour target field. This provides a unified starting point for the
target rest frame description of the nuclear dependence of a large
class of observables. For the case of the virtual $\gamma^*\to 
q\, \bar{q}$ photoabsorption cross section, we investigate then
in detail under which conditions the nuclear dependence encoded
in the Furry wavefunctions can be parametrized by a $q\, \bar{q}$
QCD dipole cross section. The important condition is colour
triviality, i.e., the property that for arbitrary $N$-fold
rescattering contributions the only non-vanishing colour trace
is $N_c\, C_F^N$. We give proofs for the colour triviality of
the inelastic, diffractive and total photoabsorption cross section 
measured inclusively or with one jet resolved in the final state.
Also, we list examples for which colour interference
effects remain. Colour triviality allows us to write the $\gamma^*\to 
q\, \bar{q}$ contribution to the DIS
nuclear structure function $F_2$ for small Bjorken $x_{\rm Bj}$ 
in terms of a path integral which describes the transverse size
evolution of the $q\, \bar{q}$ pair in the nuclear colour field. 
This expression reduces in an opacity expansion to the $N=1$
result of Nikolaev and Zakharov, and in the eikonal approximation
to the Glauber-type rescattering formulas first derived by 
Mueller. In the harmonic oscillator approximation of the
path integral, we quantify deviations from the eikonal limit.
Their onset is characterized by the scales $L/l_f$ and 
$E_\perp^{\rm tot}\, L$ which relate the longitudinal extension
$L$ of the nuclear target to the coherence length $l_f$ and the
total transverse energy $E_\perp^{\rm tot}$ accumulated by the
$q$-$\bar{q}$-pair.
\end{abstract}


\section{Introduction}
\label{sec1}

The partonic interpretation of physical processes and their
nuclear dependence is Lorentz frame dependent. Drell-Yan, e.g., 
is a $q$-$\bar{q}$ $\to \gamma^*$-fusion in the infinite momentum 
frame~\cite{BBL89}, but it becomes a $\gamma^*$-bremsstrahlung radiation off 
the projectile quark if viewed in the target rest 
frame~\cite{Boris,BHQ97}. Similarly, nuclear 
shadowing is due to the recombination of partons from different
nucleons, if viewed in the infinite momentum frame~\cite{MQ86}. 
In the target frame, it arises from non-additive contributions 
of the rescattering of the hadronic $\gamma^*$-Fock states in a 
spatially extended medium~\cite{M90,NZ91,PW99,M99,KL99}.
The descriptions in different Lorentz frames are equivalent, of course.
But depending on the physical problem at hand, a well-chosen
Lorentz frame may provide a particularly simple partonic interpretation.
To study the nuclear dependence of physical 
processes, the target rest frame provides arguably the most
intuitive picture. QCD factorization theorems can be 
expected to hold for the nuclear dependence of
high-$p_t$ processes only~\cite{LQS94,XG98,FSSM99}, and thus a large
variety of other approaches to the nuclear dependence of hard
processes exists in the literature
~\cite{BBL89,NZ91,GW94,WGP95,Z96,BDMPS2,BDMPS3,BDMS,BDMS-Zak,Z98,KST98,WG99,BDMS2,GLV99,KST99}.
These exploit that in the target rest frame, the nuclear dependence 
of a physical process $P$ with transition amplitude 
$\langle \Psi_i\vert\ P\vert \Psi_f\rangle$ can be attributed often
to the multiple rescattering of the hard in- and outgoing 
partons inside the soft spatially extended target colour field.
In the simplest cases, this rescattering effect can be described
by multiplying the free in- and outgoing wavefunctions with 
straight eikonal Wilson lines which account for the leading 
medium-induced colour rotation of projectile 
quarks~\cite{BBL89,N91}.

In the present work, we derive and study a more refined approximation
scheme for in- and outgoing wavefunctions  $\Psi_i$ and $\Psi_f$
which includes the leading transverse dynamical evolution of the 
projectile partons inside the target. To this end,
we derive in section~\ref{sec2} explicit expressions for
$\Psi_i$ and $\Psi_f$ which approximate to leading order 
$1/E$ in the norm {\it and} next to leading order in the phase 
the solution of the Dirac
equation in the presence of a spatially extended colour field.
The $1/E$-corrections included in these solutions are known to
provide the leading contribution for observables
which are essentially determined by the destructive interference 
between different production amplitudes as, e.g., the non-abelian 
Landau-Pomeranchuk-Migdal (LPM) 
effect~\cite{GW94,WGP95,Z96,BDMPS2,BDMPS3,BDMS,BDMS-Zak,Z98,KST98,WG99} 
or the nuclear dependence of Drell-Yan pair 
production~\cite{BHQ97,KST98}. Indeed, our derivation of $\Psi_i$ 
and $\Psi_f$ draws on exactly the same approximation schemes which
were used in recent studies~\cite{GW94,BDMPS1} of the nuclear 
dependence of these observables. In section~\ref{sec2}, however, we 
discuss this rescattering effect without reference to a particular 
observable. In this way, we obtain a compact expression which turns 
out to be the non-abelian generalization of the abelian Furry 
approximation~\cite{KST98,WG99} and which may serve as building 
block in the calculation of very different nuclear dependencies.

The main technical complication in working with non-abelian
Furry wavefunctions is that they involve a path-integral over a
path-ordered non-abelian Wilson line. In section~\ref{sec3} and
~\ref{sec4}, we show for the example of the $\gamma^* \to q\, \bar{q}$
photodissociation process how explicit and exact calculations can be
done despite this complication. The key step is a diagrammatic 
technique first used by Mueller and collaborators
~\cite{M90,BDMPS2,BDMPS3,BDMS,BDMS-Zak,KM98} which allows to 
establish the colour triviality of certain cross sections involving 
$N$-fold rescattering processes. Here, colour triviality means that 
the contribution to a medium-dependent observable to $N$-th order in 
the opacity involves only colour traces which reduce to the $N$-th
power of the Casimir. This renders the problem essentially abelian.
In section~\ref{sec4c}, we derive the corresponding diagrammatic 
identities for our configuration space formulation of partonic 
rescattering. We then use these identities in explicit proofs of 
the colour triviality of the inelastic and diffractive part of the inclusive 
and one-fold differential (i.e. one jet resolved) photoabsorption cross 
section. At least for the $\gamma^* \to q\, \bar{q}$ photodissociation 
cross section, all colour trivial observables turn out to be given
in terms of the transverse dynamical evolution of a QCD dipole which 
is described by a simple path integral ${\cal K}$. We discuss the 
eikonal limit in which the transverse dynamical evolution is neglected,
and we quantify the leading corrections to this limiting case.
 
We choose for all explicit calculations the photodissociation process  
$\gamma^* \to q\, \bar{q}$ mainly since it is the simplest example which 
allows to illustrate the main technical difficulties associated with the 
non-abelian Furry approximation and colour interference effects. 
The $\gamma^* \to q\, \bar{q}$ process is a quantitatively 
significant contribution to the nuclear structure function $F_2$, 
but phenomenological applications require the inclusion of
processes with initial gluon radiation (e.g. $\gamma^* \to q\, \bar{q}\, g$).
These are known to contribute in the aligned jet region, where the
transverse $q\, \bar{q}$ separation is large, to leading
order in $1/ Q^2$. Discussion of these radiative contributions
involves the quark-gluon vertex and complicates the analysis
of colour interference effects significantly. This lies beyond
the scope of a first illustration of the non-abelian Furry
approximation, given here. We shall address the corresponding
additional technical problems in a subsequent work where we
study colour interference in the non-abelian LPM-effect. In the
present paper, we discuss only two aspects of the very general
question to what extent our approach can be applied to other 
or more exclusive observables or to other models of the medium: First,
we give in section~\ref{sec4b} and ~\ref{sec4e} examples of 
(more exclusive) photoabsorption processes which are not colour 
trivial, thus indicating the limitations of the approach advocated here. 
Second, we argue in section~\ref{sec5} that the specific model
ansatz for in-medium rescattering is not instrumental for our
results. In the Conclusions, we summarize our main results and 
we shortly comment on further perspectives.

\section{Asymptotic wavefunctions for rescattering particles}
\label{sec2}

\begin{figure}[h]\epsfxsize=8.7cm 
\centerline{\epsfbox{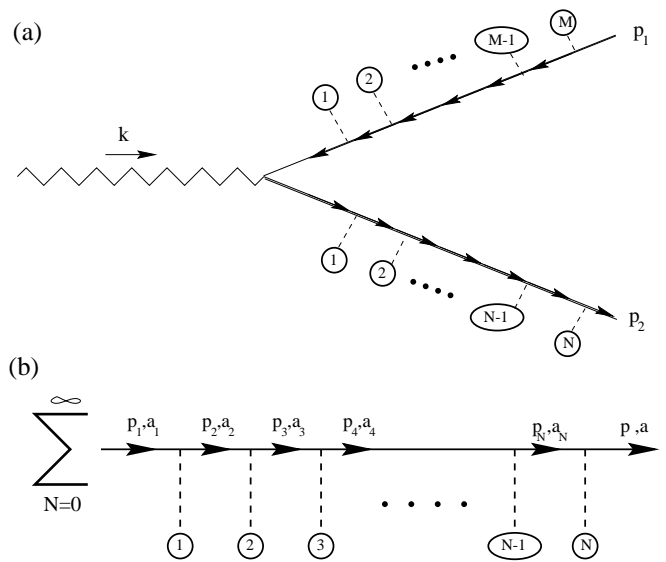}}
\vspace{0.5cm}
\caption{(a) Contribution to the photodissociation process 
$\gamma^*\to q\, \bar{q}$ with $N$- and $M$-fold final state
rescattering for the outgoing quark and antiquark.
(b) Sum over rescattering diagrams for an outgoing antiquark which
leaves the production vertex with momentum $p_1$ and colour $a_1$. 
To leading order in opacity, this resums all rescattering effects and 
defines the non-abelian Furry wavefunction (\protect\ref{2.12}).
}\label{fig1}
\end{figure}

We want to calculate the nuclear dependence of some physical 
process $P$ with transition amplitude 
$\langle \Psi_i\vert\ P\vert \Psi_f\rangle$ by describing how
the medium affects the propagation of the in- and outgoing
wavefunctions $\Psi_i$, $\Psi_f$. For the case of 
the photodissociation process $\gamma^*\to q\bar{q}$ depicted 
in Fig.~\ref{fig1}a this means that we write the corresponding transition 
amplitude in the form
\begin{eqnarray}
  \langle \Psi_i\vert\ P\vert \Psi_f\rangle
  &=& ie\, \int
  d^4x\,{\Psi_u}^{\dagger}(x,p_1)\, \gamma^0
  \nonumber \\
  && \qquad \qquad \times 
  \, \epsilon_\mu\, \gamma^\mu\,
  e^{i\, k\cdot x}\,
  \Psi_v(x,p_2)\ .
  \label{2.1}
\end{eqnarray}
Here, the physical process $P$ is determined by the photon-
quark vertex $ie \epsilon\cdot\gamma\, e^{ikx}$. 
In this section, we derive explicit expressions for the
medium-dependence of these asymptotic wavefunctions
irrespective of a particular process. Section
~\ref{sec2a} explains the solution of the corresponding 
abelian problem, section~\ref{sec2b} establishes the
non-abelian extension. In the remaining sections of this paper, we shall
discuss then for the example of the photodissociation (\ref{2.1})
how explicit calculations can be done. 

%
\subsection{Abelian Furry approximation}
\label{sec2a}

We consider a relativistic positron with momentum $(E_2,{\bf p}_2)$
($E_2 \gg m$). Prior to its detection, this positron undergoes multiple
small-angle scattering in a spatially extended medium, described 
e.g. by a collection $U({\bf x})$ of single scattering potentials 
$\varphi_i({\bf x})= \varphi({\bf x}-\check{\bf x}_i)$ localized 
at spatial positions $\check{\bf x}_i$,
\begin{eqnarray}
  U({\bf x}) = \sum_{i=1}^\infty\, 
  \varphi({\bf x}-\check{\bf x}_i)\, .
  \label{2.2}
\end{eqnarray}
The asymptotic electron wavefunction $\Psi$ is a solution to 
the Dirac equation in this spatially extended field:
\begin{equation}
  \left[i\, {\partial\over \partial t} - U({\bf x}) - m\,\gamma_0 +
  i\,\bbox{\alpha}\cdot\bbox{\nabla}\right]\,
  \Psi(x,p_2) = 0\, .
  \label{2.3}  
\end{equation}
Which approximation for $\Psi(x,p_2)$ keeps 
the leading medium-dependence ? An expansion of $\Psi$ in powers 
of the coupling constant does not since it amounts to an expansion 
in powers of the single scattering potential $\varphi$
rather than containing the leading order effect of $U$.
In QED, the leading $U$-dependence is kept in the Furry 
approximation~\cite{KST98,WG99} 
which is a high energy expansion of the solution of the Dirac 
equation. For the outgoing positron 
wavefunction, it reads
 \begin{eqnarray}
  \Psi_F(x,p_2) &=& e^{iE_2t - ip_2z}\,
   \hat D_2\,F({\bf x},{\bf p}_2)\,
   v({\bf p}_2)\, .
   \label{2.4}
 \end{eqnarray}
which is exact to order $O(U/E)$ and $O(1/E^2)$. Here, $\hat D_i$ 
denotes a differential operator
 \begin{eqnarray}
 \hat D_i &=&
  1- i\,\frac{\bbox{\alpha}\cdot\bbox{\nabla}}{2\,E_i} 
   - \frac{\bbox{\alpha}\cdot({\bf p}_i-{\bf n}\,p_i)}{2\,E_i}\, ,
  \nonumber\\
   && \bbox{\alpha} = \gamma_0\, \bbox{\gamma}\quad ;\quad
   z = \bbox{n}\cdot\bbox{x}\quad ;\quad
   p_i = |\bbox{p}_i|\, ,
   \label{2.5}
\end{eqnarray}
and the unit vector $\bbox{n}$ specifies the {\it longitudinal}
direction. The differential operator acts on the transverse
wavefunctions $F$. For very late times, i.e., for far 
forward longitudinal distances $x_L$, this wavefunction 
satisfies plane wave boundary conditions
\begin{eqnarray}
  F_\infty({\bf x}_\perp,x_L,{\bf p}_2) &=& 
  \exp\left\{- i\,{\bf p}_2^\perp\cdot{\bf x}_\perp
  + i\, { {{\bf p}_2^\perp}^2\over 2\, p_2} x_L\right\}\, ,
  \label{2.6} \\
  F({\bf y}_\perp,y_L,{\bf p}_2) &=& 
  \int d{\bf x}_\perp\, G({\bf y}_\perp,y_L;{\bf x}_\perp,x_L|p_2)
  \nonumber \\
  && \quad \times\, F_\infty({\bf x}_\perp,x_L,{\bf p}_2)\, .
  \label{2.7}
\end{eqnarray}
Its evolution to finite longitudinal distances is determined by the
retarded Green's function $G$ whose path integral representation
reads (for $z' > z$)
\begin{eqnarray}
  &&\qquad\qquad G(\bbox{r}, z; \bbox{r}',z'|\,p) =  \nonumber \\
  &&\int {\cal D}\bbox{r}(\xi)\,
  \exp\left\{ \int\limits_{z}^{z'}{\it d}\xi\,
  \left[\frac{ip}{2} \dot{\bf r}^2(\xi) -
  i\,U\bigl({\bf r}(\xi),\xi\bigr)\right] \right\}\, .
  \label{2.8}
\end{eqnarray}
Here, $\dot{r} = dr/d\xi$ and $G$ satisfies the 
boundary conditions ${\bf r}(z)={\bf r}$, ${\bf r}(z')={\bf r}'$, with
$G({\bf r},z;{\bf r}',z'=z|{\bf p}) = \delta({\bf r}'-{\bf r})$. The Green's
function (\ref{2.8}) describes the Brownian motion of the projectile 
particle in the plane transverse to the beam. This is the leading 
medium effect on the propagation of the projectile. Starting from 
the abelian Furry wavefunction (\ref{2.4}), the 
KST-formalism~\cite{KST98,WG99} 
then allows to describe e.g. the medium dependence of the  
LPM-bremsstrahlung spectrum~\cite{KST98,WG99}. One of the main motivations
for what follows is the question to what extent the application of
the same approach to non-abelian problems is justified.

\subsection{Non-abelian Furry approximation}
\label{sec2b}

To find the non-abelian generalization of the Furry approximation, we 
consider a spatially extended static colour potential of the form
\begin{eqnarray}
  A_\mu({\bf x}) &=&
  \delta_{0\mu}\, \sum_{i=1}^\infty\, \varphi^a_i({\bf x})\, T^a\, ,
  \label{2.9} \\
  \varphi^a_i({\bf x}) &=& \varphi({\bf x}-\check{\bf x}_i)\, 
  \delta^{a\, a_i}\, ,
  \label{2.10}
\end{eqnarray}
where $T^{a}$, $(a=1, \dots, N_c^2-1)$, denote the generators of the 
$SU(N_c)$ colour representation of the projectile parton. The $i$-th 
scattering center is located at $\check{\bf x}_i$ and exchanges a 
specific colour charge $a=a_i$.

For the spatial support of the potentials $\varphi^a_i$ we take the
rapid fall-off of a Yukawa potential with some Debye screening mass 
$M$. We assume that the mean free path of the projectile parton in 
the medium is taken to be much larger than $1/M$. This ansatz for
(\ref{2.10}) is known as Gyulassy-Wang model~\cite{GW94,WGP95} and 
was originally introduced to mimic rescattering effects of hard partons 
in the colour-deconfined matter created in the early phase of 
a relativistic heavy ion collision. 

To leading order $O(1/E)$, the Feynman diagram in Fig.~\ref{fig1}b is 
the only $\alpha_s^N$ rescattering term. Rescattering contributions 
involving 3- or 4-gluon vertices are known to come with spatial
suppression factors in the Gyulassy-Wang model~\cite{GW94,BDMPS3}.
The $N$-fold rescattering of a hard parton in the 
colour field (\ref{2.9}) is thus determined by
\begin{eqnarray}
  I^{(N)}({\bf y}) &=&
  e^{-i\, {\bf p}_1\cdot {\bf y}}\,
 {\cal P}\, \left( 
  \prod_{i=1}^N\int {d^3{\bf p}_i\over (2\pi)^3}\, d^3{\bf x}_i\,
                    {i\, (\not{p}_i + m)\, \gamma_0\over
                     {p_i^2 - m^2 + i\,\epsilon}} \right. 
                   \nonumber \\
                   && \times
                    \left.
                    \lbrack -i\, A_0({\bf x}_i)\rbrack
                    \, e^{-i\, {\bf x}_i\cdot 
                      ({\bf p}_{i+1} - {\bf p}_i)} \right)\,
                    v^{(r)}({\bf p})\, .
  \label{2.11}
\end{eqnarray}
Here, the quark leaves some production vertex with momentum $p_1$ and 
colour $a_1$. It undergoes $N$ gluon exchanges with the spatially extended
colour potential $A_\mu$, emerging in the  ``final state'' with 
momentum $p$ and colour $a$.  $I^{(N)}({\bf y})$ is a 
matrix in the colour representation of the projectile. 
The path-ordering ${\cal P}$ implies that $A_0({\bf x}_{i+1})$
stands to the right of $A_0({\bf x}_i)$. The diagram
in Fig.~\ref{fig1} denotes the component $I^{(N)}_{a_1\, a}$ of this
matrix. 
The momentum transfers to the quark line are written as Fourier
transforms of the static scattering potential with respect to the
relative momenta ${\bf p}_{i+1} - {\bf p}_i$. Finally, instead of 
an explicit production vertex, we have introduced in (\ref{2.11}) the 
incoming plane wave $\exp\left[-i\, {\bf p}_{1} \cdot {\bf y}\right]$. 
This factor allows to glue the above Feynman diagram via 
${\bf y}$-integration onto another subprocesses $P$ without 
specifying $P$ at the present stage.

In appendix~\ref{appa}, we approximate the norm of $I^{(N)}$ to
leading order $O(1/E)$ and the phase to next to leading order.
Summing then over contributions for arbitrarily many $N$ 
rescatterings in the medium, we find the non-abelian Furry 
wavefunction
\begin{eqnarray}
  {\Psi_v}(y^0,{\bf y},p) &=& e^{i\, E\, y^0}
                \sum_{N=0}^\infty\, I^{(N)}({\bf y})
  \nonumber \\
  &=& e^{i\, E\, y^0 - i\, p\, y_L}\, {\hat D}\,
  F({\bf y},{\bf p})\, v^{(r)}({\bf p})\, . 
  \label{2.12} 
\end{eqnarray}
This expression compares directly to the abelian Furry approximation
(\ref{2.4}). In (\ref{2.12}), 
${\hat D}$ is the operator (\ref{2.5}) with $p_i = p$, and
$F$ is the outgoing transverse wavefunction evolved from its
asymptotic plane wave form to ${\bf y}$ with the Green's function
$\bar G$,
\begin{eqnarray}
  F({\bf y},{\bf p}) &=& \int d{\bf x}_\perp\, 
 \bar G({\bf y}_\perp,y_L;{\bf x}_\perp,x_L\vert p) 
  \nonumber \\
  && \times
  F_\infty({\bf x}_\perp,x_L, {\bf p})\, .
  \label{2.13}
\end{eqnarray} 
The Green's function $\bar{G}$ is the non-abelian generalization
of the Green's function (\ref{2.8}). It can be defined 
explicitly by its expansion in powers of the scattering potential $A_0$:
\begin{eqnarray}
  &&\bar G({\bf r},z;{\bf r}',z'|p) \equiv  G_0({\bf r},z;{\bf r}',z'|p) 
  -i\int\limits_{z}^{z'} d\xi
  \nonumber \\
  && \qquad \times 
  \int d{\bbox \rho}\, G_0({\bf r},z;{\bbox \rho},\xi|p)\, 
  A_0({\bbox \rho},\xi)\, 
  G_0({\bbox \rho},\xi;{\bf r}',z'|p)\,
  \nonumber \\
  && \quad + {\cal P} \int\limits_{z_L}^{x_L} d\xi_1\, 
       \int\limits_{\xi_1}^{x_L} d\xi_2\, 
       \int d{\bbox \rho}_1\, d{\bbox \rho}_2\,
       G_0({\bf r},z;{\bbox \rho}_1,\xi_1|p)
  \nonumber \\
  && \qquad \times  i\, 
  A_0({\bbox \rho}_1,\xi_1)\, 
  G_0({\bbox \rho}_1,\xi_1;{\bbox \rho}_2,\xi_2|p)\, 
  \nonumber \\
  && \qquad \times i\, 
  A_0({\bbox \rho}_2,\xi_2)\, 
  \bar{G}({\bbox \rho}_2,\xi_2;{\bf r}',z'|p)\, .
  \label{2.14}
\end{eqnarray}
Here, $G_0$ is the free non-interacting Green's function
\begin{equation}
  G_0({\bf r},z;{\bf r}',z'|p) \equiv \frac{p}{2\pi i(z'-z)}
  \exp\left\{ \frac{ip\, \left({\bf r}-{\bf r}'\right)^2}{2(z'-z)} 
              \right\}\, ,
  \label{2.15}
\end{equation}
and the path ordering ${\cal P}$ in (\ref{2.14}) ensures that the potential
$A_0({\bbox \rho}_2,\xi_2)$ stands to the right of the potential
$A_0({\bbox \rho}_1,\xi_1)$. We shall use the Green's function 
$\bar G$ in (\ref{2.14}) for $z'>z$ and $z>z'$ by defining
\begin{equation}
  \bar G({\bf r}',z';{\bf r},z|p) \equiv 
  \bar G^\dagger({\bf r},z;{\bf r}',z'|p)\, ,\qquad
  \hbox{for $z'>z$}\, .
  \label{2.16}
\end{equation}
For a hermitian scattering potential, this definition is compatible with 
the representation (\ref{2.14}). 
Hermitian
conjugation automatically inverts the path-ordering. A very compact
notation for the expansion (\ref{2.14}) can be given in terms of
a path-ordered Wilson line $W\bigl([{\bf r}];z,z'\bigr)$
which follows the non-abelian potential $A_0$ from initial 
position $({\bf r}(z),z)$ to final position $({\bf r}(z'),z')$
along the path ${\bf r}(\xi)$
\begin{eqnarray}
  &&\bar G({\bf r},z;{\bf r}',z'\vert p) 
  =  \nonumber \\
  &&=\int {\cal D}\bbox{r}(\xi)\,
  \exp\left\{ \frac{ip}{2} \int\limits_{z}^{z'}{\it d}\xi\,
  \dot{\bf r}^2(\xi)\right\}\, 
  W\bigl([{\bf r}];z,z'\bigr) \, ,
  \label{2.17}\\
&&  W\bigl([{\bf r}];z,z'\bigr) = {\cal P}\,
  \exp\left\{ -i\, \int\limits_z^{z'}{\it d}\xi\,
              A_0({\bf r}(\xi),\xi) \right\}\, .
  \label{2.18}
\end{eqnarray}
Again, operators at larger longitudinal distances ($z'>z$) stand
to the right. Expanding the Wilson line to fixed order 
$O(A_0^n)$ coincides with the representation (\ref{2.14}). 
The non-abelian Furry approximation thus differs from the abelian one 
essentially by path-ordering in the Green's function $\bar{G}$ which 
describes the dynamical evolution of the transverse wavefunction.

To obtain the result (\ref{2.12}), it is crucial that we approximate 
first the $N$-fold rescattering diagram $I^{(N)}$, keeping the phase
factor to order $O(1/E)$ and then resum contributions for 
arbitrary $N$. If one parallels the 
derivation of the abelian case by directly starting from the 
solution of the non-abelian Dirac equation in the presence 
of a spatially extended colour field, one arrives at the recursive
solution of Buchm\"uller and Hebecker~\cite{BH96}. The
starting point for their recursion formula is a straight eikonal 
Wilson line, and the recursion includes corrections to fixed order 
in energy. An explicit $n$-fold iterated expression is thus 
correct to fixed order $O(1/E^n)$, but it does not contain
the $O(1/E)$-corrections to the {\it phase} of the wavefunction,
which is characteristic for the Furry approximation. Our diagrammatic
approach allows to keep this phase. For large but finite incident
energy, our solution is thus characteristically different from that
given in Ref.~\cite{BH96}. [We have checked, e.g., that the abelian
version of ~\cite{BH96} does not lead to the correct medium dependence
of the QED bremsstrahlung spectrum while the abelian version of 
(\ref{2.12}) does. This difference  of both solutions is expected
since the bremsstrahlung spectrum is sensitive to interference effects
which stem from the $O(1/E)$-contributions to the phase of the
wavefunctions~\cite{WG99}.]

In the infinite energy limit, the solution (\ref{2.12}) reduces 
to the straight eikonal Wilson line
\begin{eqnarray}
  &&\lim_{\nu\to\infty} 
    \bar{G}({\bf r},z;{\bf x}_\perp,x_L|\nu)
  = W([{\bf r}_s];z,x_L)\, ,
  \label{2.19}\\
  &&{\bf r}_s(\xi) = {\bf x}_\perp \frac{\xi-z}{x_L-z}
    + {\bf r} \frac{x_L-\xi}{x_L-z}\, ,
  \label{2.20}
\end{eqnarray}
which is the leading order input for the recursion formula of
Ref.~\cite{BH96}. This expression is often used to describe 
the leading colour rotation of the hard parton in a soft 
colour field~\cite{BBL89,N91}.
 
The Furry wavefunction (\ref{2.12}) derived here allows for the
description of the nuclear dependence of a large class of physical
observables. To discuss its limitations, we note that
QCD factorization theorems for 
observables including rescattering effects have been established 
(and can be expected to hold) only for processes for which 
kinematical constraints ensure very high transverse 
momentum transfers ($\gg \Lambda_{QCD}$),~\cite{LQS94,XG98,FSSM99}. 
The description of {\it soft} nuclear rescattering 
of hard partons always depends on additional assumptions and different
approaches may be taken~\cite{BBL89,GW94,BDMPS2}. In particular, we assume
(as all other treatments do) that the leading nuclear dependence
can be obtained using the idealization of ``asymptotic'' parton 
wavefunctions, i.e., without considering parton fragmentation.
Since parton fragmentation and soft parton rescattering involve
transverse momenta of the same order, this is a requirement on
the spatial separation of the two phenomena and amounts to an 
energy-dependent limit
on the longitudinal extension $L$ of the nuclear target up to
which this description can be expected to apply. Moreover, the
ad hoc separation of the transition amplitude into a production $P$ 
and a final state wavefunction including rescattering effects may
be oversimplified for some processes. We think, e.g., of the 
production and rescattering of a heavy quarkonium state with
specific quantum numbers. Taking $P$ to be the production of
the heavy $q$-$\bar{q}$-pair, it is not clear a priori to what 
extent final state rescattering modifies the quantum numbers of
this state. First studies indicate~\cite{PH98} that the discussion of this
problem requires a classification of the hardness of medium-induced
momentum transfers which lies outside the scope of the present
calculation. To sum up: the discussion of the nuclear dependence
of hard observables  in terms of $\langle \Psi_i| P|\Psi_f\rangle$
seems justified if no other soft scale (introduced e.g. by parton
fragmentation or by the binding energy of the final state) interfers
significantly with the final state rescattering effect described
by the Furry wavefunctions. The non-abelian 
LPM-effect~\cite{GW94,WGP95,BDMPS2,BDMPS3,BDMS,BDMS-Zak}, the
nuclear dependence of Drell-Yan yields~\cite{BHQ97,KST98} and 
nuclear shadowing~\cite{NZ91,PW99} are prominent examples for 
which a description in terms of  
$\langle \Psi_i| P|\Psi_f\rangle$ seems suitable.

\section{Photodissociation}
\label{sec3}

\begin{figure}[h]\epsfxsize=6.7cm 
\centerline{\epsfbox{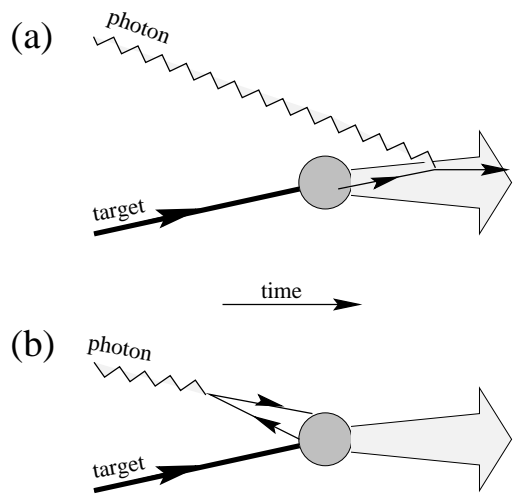}}
\vspace{0.5cm}
\caption{Two possible time orderings for the interaction of
a virtual photon with a nuclear target. At small Bjorken $x$,
the $\gamma^*\to q\bar{q}$ photodissociation process (b)
dominates over the $\gamma^*$-$q$-fusion, see 
Ref.~\protect\cite{PW99}.
}\label{fig2}
\end{figure}

In this section, we give a closed expression for the 
$\gamma^*\to q\, \bar{q}$  photodissociation cross section
$\sigma^{\gamma^*\to q\bar{q}}$ in terms of the non-abelian
Green's function $\bar{G}$ and the squared ingoing wavefunction
$\Phi$ of a freely evolving $q$-$\bar{q}$-pair. This 
photodissociation cross section contributes to 
the total virtual photoabsorption cross section
$\sigma_{total}^{\gamma^*}$ which is related to the deep 
inelastic structure function $F_2$,~\cite{F89}
\begin{equation}
  F_2(x,Q^2) = \frac{Q^2}{4\pi^2\, \alpha_{\rm em}}
               \sigma_{total}^{\gamma^*}(x,Q^2)\, .
  \label{3.1}
\end{equation}
In the target rest frame, one can show that photodissociation
as depicted in Fig.~\ref{fig2}b dominates for small Bjorken $x$ 
over the $\gamma^*$-$q$ fusion process shown in Fig.~\ref{fig2}a
~\cite{PW99}. The leading corrections to 
$\sigma_{\rm total}^{\gamma^*\to q\bar{q}}$ come from initial state gluon
radiation~\cite{PW99,BGH99} which we neglect in what follows.

We consider a virtual photon of four momentum
$k = (\nu,{\bf 0}_\perp,\sqrt{\nu^2 + Q^2})$ which dissociates 
into a quark-antiquark pair in the time interval $-T/2 < t < T/2$. 
The dissociation cross section is
\begin{equation}
  \sigma_{\rm total}^{\gamma^*\to q\bar{q}} = \frac{1}{2\, |k^L|\, T}
  \int dX\, |S_{fi}|^2\, ,
  \label{3.2}
\end{equation}
where the phase space element is written in terms of the on-shell
momenta of the outgoing quarks,
\begin{equation}
  dX = \frac{d^3p_1}{(2\pi)^3\, 2\, |E_1|}\,
        \frac{d^3p_2}{(2\pi)^3\, 2\, |E_2|}\, .
  \label{3.3}
\end{equation}
The rescattering of these quarks in the spatially extended
colour field is described by the Furry wavefunctions in the 
dissociation amplitude
\begin{eqnarray}
  S_{fi} &\equiv& S_{fi}^\mu\, \epsilon_\mu\, ,
  \label{3.4}\\
  S_{fi}^\mu &=& ie\, \int
  d^4x\,{\Psi_u}^{\dagger}(x,p_1)\,
  \gamma^0\, \gamma^\mu\,
  \nonumber \\
  && \quad \times e^{-\epsilon\, |z|}\,
  e^{i\, k\cdot x}\,
  \Psi_v(x,p_2)\ .
  \label{3.5}
\end{eqnarray}
To discuss the case of different photon polarizations, we  
write this disscociation amplitude as a contraction of
$S_{fi}^\mu$ with the polarization vector $\epsilon_\mu$.
To shorten our notation, we do not keep track of the fractional
charge $e_q$ of the quark  and of the number of flavours. To 
do this, all our final results have to be supplemented by a
sum over the available flavour channels weighted by $e_q^2$.

%
Inserting the Furry wavefunctions for quark and antiquark, 
see (\ref{2.12}), we can separate in (\ref{3.5}) an energy 
conserving $\delta$-function, $S_{fi}^\mu = ie\, (2\pi)\, 
\delta(E_2-E_1-\nu)\, M_{fi}^\mu$. We introduce
the fraction $\alpha$ of the initial energy carried by one quark,
$E_2 = \alpha\, \nu$, $E_1 = -(1-\alpha)\, \nu$. Here, $E_1$ is negative,
since the corresponding 4-momentum $p_1$ flows into the quark-photon
vertex, see Fig.~\ref{fig1}a. After coordinate
transformation $(p_1^L,p_2^L) \to (\alpha,k^L)$, the 
photodissociation cross section reads
\begin{eqnarray}
  \sigma_{\rm total}^{\gamma^*\to q\bar{q}} &=& 
  \alpha_{\rm em} \int \frac{d\alpha}{4\, \nu^2\, \alpha\, (1-\alpha)}
  \nonumber \\
  && \times \int \frac{d{\bf p}_1^\perp}{(2\pi)^2}\, 
  \frac{d{\bf p}_2^\perp}{(2\pi)^2}\, 
  \langle|M_{fi}|^2\rangle \, .
  \label{3.6}
\end{eqnarray}
Here, we have used Fermi's golden rule and the $k^L$-integration to
eliminate the two energy-conserving $\delta$-functions. The
brackets $\langle \dots\rangle$ denote an in-medium average over
the colour field $A_\mu$ which will be specified below. The probability
$|M_{fi}|^2$ can be written in terms of the full interacting Green's
functions (\ref{2.17}) as
%
%
\begin{eqnarray}
 &&|M_{fi}|^2 = 
      \int {\it d}^3{\bf y}\, {\it d}^3\bar{\bf y}\, 
      e^{i\, q\, (y_L-\bar{y}_L)}\, 
      e^{-\epsilon(|y_L|+|\bar{y}_L|)} \int 
      {\it d}{\bf x}_\perp
      \nonumber \\
 && \quad \times {\it d}{\bf x'}_\perp\, 
      {\it d}{\bf \bar x}_\perp\, 
      {\it d}{\bf \bar x'}_\perp\, 
      e^{i\, {\bf p}_1^\perp\cdot(
                {\bf x'}_\perp - {\bf \bar x'}_\perp)}\, 
      e^{-i\, {\bf p}_2^\perp\cdot(
                {\bf x}_\perp - {\bf \bar x}_\perp)}\, 
      \nonumber \\
  &&\quad \times 
    \bar G({\bf x};{\bf y}\vert p_2)\, 
    \epsilon_\mu\hat\Gamma^\mu\, 
    \bar G({\bf y};{\bf x'}\vert p_1)
    \nonumber \\
  &&\quad \times 
    \bar G({\bf \bar x'};{\bf \bar y}\vert p_1)\, 
    \epsilon_\nu {\hat{\Gamma^*}}^\nu\, 
    \bar G({\bf \bar y};{\bf \bar x}\vert p_2)\, .
  \label{3.7}
\end{eqnarray}
In general, all spatial coordinates can be different for $M_{fi}$ 
and $M_{fi}^\dagger$. We characterize those for $M_{fi}^\dagger$ 
by a bar. The spinor structure of the Furry wavefunctions (\ref{2.12})
is contained in (\ref{3.7}) in the vertex function
\begin{equation}
 \widehat\Gamma^\mu =
 {u^{(r')}}^{\dagger}(-p_1)\,\hat D^*_1\,
 \gamma^0\, \gamma^\mu\,\hat D_2\,v^{(r)}(p_2)\, .
 \label{3.8}
\end{equation}
In appendix~\ref{appb}, we discuss how these vertex functions 
combine with free Green's functions $\bar G_0$ to the incoming 
$q$-$\bar{q}$ Fock state. In terms of the square 
$\Phi(\Delta {\bf z}; \Delta {\bf \bar z};\alpha)$ of
this Fock state, the differential photodissociation
cross section (\ref{3.6}) takes the form
\begin{eqnarray}
 && \frac{\sigma_{\rm total}^{\gamma^*\to q\bar{q}}}{d\alpha\,
   d{\bf p}_1^\perp\, d{\bf p}_2^\perp}
 \nonumber \\
 && =  \frac{\alpha_{\rm em}}{(2\pi)^4}
    \int {\it d}{\bf b}_1\,{\it d}{\bf b}_2\, 
    {\it d}{\bf \bar b}_1\, {\it d}{\bf \bar b}_2\,
    \Phi(\Delta {\bf b}; \Delta {\bf \bar b};\alpha)
      \nonumber \\
 && \quad \times 
    \int {\it d}{\bf x}_\perp\, {\it d}{\bf x'}_\perp\, 
      {\it d}{\bf \bar x}_\perp\, 
      {\it d}{\bf \bar x'}_\perp\, 
      e^{i\, {\bf p}_1^\perp\cdot(
                {\bf x'}_\perp - {\bf \bar x'}_\perp)}\, 
      e^{-i\, {\bf p}_2^\perp\cdot(
                {\bf x}_\perp - {\bf \bar x}_\perp)}
      \nonumber \\
  &&\quad \times 
    \Big\langle\, \bar G({\bf \bar b}_2;{\bf \bar x}\vert p_2)
    \bar G({\bf x};{\bf b}_2\vert p_2)
      \nonumber \\
  &&\quad \times 
    \bar G({\bf b}_1;{\bf x'}\vert p_1)\,
    \bar G({\bf \bar x'};{\bf \bar b}_1\vert p_1)\Big\rangle\, ,
  \label{3.9}
\end{eqnarray}
where $\Delta {\bf b} = {\bf b}_1 - {\bf b}_2$ and 
$\Delta \bar {\bf b} = \bar {\bf b}_1 - \bar {\bf b}_2$.
The explicit form of the squared incoming wavefunction $\Phi$
is derived in appendix~\ref{appb}.

\begin{figure}[h]\epsfxsize=8.7cm 
\centerline{\epsfbox{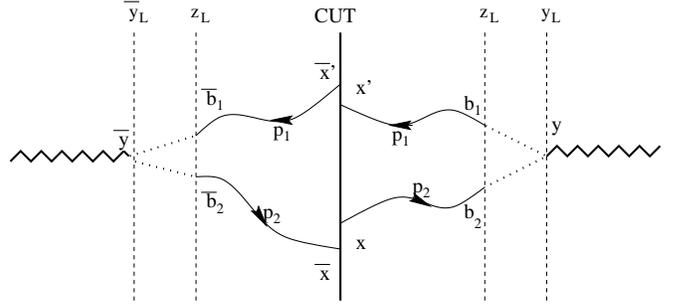}}
\vspace{0.5cm}
\caption{Representation of the $\gamma^* \to q\, \bar{q}$ 
photodissociation cross section (\protect{\ref{3.6}}), (\protect{\ref{3.7}}). 
In the amplitude $M_{fi}$, the photon dissociates at position $({\bf y},y_L)$
into a quark and an antiquark. These propagate with free non-interacting
Green's functions (dotted lines) up to the front end of the 
target at longitudinal position $z_L$. From $z_L$ onwards, the quark
and antiquark are propagated inside the colour field with interacting
Green's functions (full lines) up to their asympotic on-shell states
(``cut''). For the complex conjugate amplitude, the analogous evolution
of the hadronic $q\bar{q}$ Fock state is depicted on the other side of
the cut.
}\label{fig3}
\end{figure}

In equation (\ref{3.9}), we assume that the nuclear target has
nonvanishing density only for some longitudinal positions $z_L > 0$.
The integration variables ${\bf b}_i$, ${\bf \bar b}_i$ denote
the transverse boundary values of the Green's functions at the
front end $z_L = 0$ of the nuclear target. Fig.~\ref{fig3} gives
a graphical representation for (\ref{3.9}) which we shall heavily
use in what follows. According to Fig.~\ref{fig3}, the virtual 
photon wavefunction starts interacting with the nuclear medium at 
$z_L=0$ in both the amplitude and complex
conjugate amplitude {\it after} dissociating at longitudinal 
positions $y_L$, $\bar{y}_L$ with  $y_L,\bar{y}_L < 0$.
We emphasize, however, that despite appearance, equation (\ref{3.9}) 
as well as its graphical representation in Fig.~\ref{fig3} also 
include the case that
the photon vertex dissociates inside the target at $y_L > 0$. Technical
details of how this is handled are given in 
appendix~\ref{appb}.
%

\section{Opacity expansion and colour triviality}
\label{sec4}
In this section, we analyze the photoabsorption cross section
(\ref{3.9}) by expanding the interacting Green's functions
$\bar{G}$ in powers of the scattering potential $A_0({\bf x})$.
In a first step, we focus on the total photoabsorption cross section
\begin{eqnarray}
  \sigma_{total}^{\gamma^*\to q\, \bar{q}}
  &=& \alpha_{\rm em}\, \int d\alpha\, 
    \int {\it d}{\bf b}_1\, {\it d}{\bf b}_2\, 
    {\it d}{\bf \bar b}_1\, {\it d}{\bf \bar b}_2\,
    \Phi(\Delta {\bf b}; \Delta {\bf \bar b};\alpha)
    \nonumber \\
 && \times \Big\langle
      \int {\it d}{\bf x}_\perp\, 
      \bar G({\bf \bar b}_2;{\bf x}\vert p_2)\,
      \bar G({\bf x};{\bf b}_2\vert p_2)
    \nonumber \\
 && \times \int 
      {\it d}{\bf x'}_\perp\, 
    \bar G({\bf b}_1;{\bf x'}\vert p_1)\,
    \bar G({\bf x'};{\bf \bar b}_1\vert p_1) \Big\rangle\, .
  \label{4.1}
\end{eqnarray}   
To make sense of an expansion in powers of $A_0$, we have to
specify the in medium average $\langle \dots \rangle$ 
in terms of $A_0$. To this end, we specify for the colour
potential (\ref{2.9}) the contribution of a single scattering 
potential, centered at $(\check{\bf r}_i, \check{z}_i)$, 
\begin{eqnarray}
  \varphi_i^a({\bf x}_\perp,\xi) &=& \delta^{a\, a_i}
  \int \frac{d^3\bbox{\kappa}}{(2\pi)^3}\,
  \nonumber \\
  && \qquad \times 
  a_0(\bbox{\kappa}_\perp)\, 
  e^{-i({\bf x}_\perp - \check{\bf r}_i)\cdot \bbox{\kappa}_\perp}
  e^{-i(\xi-\check{z}_i)\kappa^L}\, .
  \label{4.2}
\end{eqnarray}
Here, we have approximated the argument of the single scattering 
potential $a_0(\bbox{\kappa}) \approx a_0(\bbox{\kappa}_\perp)$.
This implies that the momentum transfer occurs at a fixed 
longitudinal position $\xi=\check{z}_i$. It is the standard 
approximation in the high energy limit where the dominant momentum 
transfer from the medium is transverse, and it motivates the use
of time-ordered perturbation theory to rescattering 
problems~\cite{GW94,BDMPS2}.

Starting from (\ref{4.2}), we define the medium average 
$\langle \dots \rangle$ as an average over the transverse 
and longitudinal positions $(\check{\bf r}_i, \check{z}_i)$ 
of the scattering potentials and the colour factors $a_i$,
\begin{eqnarray}
  \langle f \rangle &\equiv& \frac{1}{A_\perp}\, 
  \left(\prod_{i=1}^N \sum_{a_i}\, \int d\check{\bf r}_i\, 
    d\check{z}_i\right)\,
  \nonumber \\
  && \qquad \times f(\check{\bf r}_1,\dots,\check{\bf r}_N;
    \check{z}_1,\dots,\check{z}_N; a_1,\dots,a_N)\, .
  \label{4.4}
\end{eqnarray}
Here, $A_\perp$ is a total transverse area which we divide out 
to regain the cross section per unit transverse area.
$N$ is the number of different single scattering potentials
up to which the function $f$ is expanded. 

To simplify notation, we replace in what follows the discrete 
sum over $\check{z}_i$ in (\ref{2.9}) by an integral over the
density $n$ of scattering centers
\begin{equation}
  A_0({\bf x}_\perp,\xi) = \int d\check{z}_i\, n(\check{z}_i)\,
  \varphi_i^a({\bf x}_\perp,\xi)\, T^a\, .
  \label{4.3}
\end{equation}
Since the effective momentum transfer from the single potential
$\varphi_i^a({\bf x}_\perp,\xi)$ occurs at $\xi=\check{z}_i$, we
shall often work with $\xi$ as integration variable. 
The expansion of the photoabsorption cross section to $N$-th order
in $A_0$ is an expansion in the $N$-th order of the 
opacity parameter $\alpha_s^2\int d\xi\, n(\xi)$. In what follows,
we study the perturbative expansion in this parameter.

\subsection{N=1 result of Nikolaev and Zakharov}
\label{sec4a}
As a first illustration of the above formalism, we expand the
integrand of the total photoabsorption cross section (\ref{4.1})
to first order in the opacity parameter $\alpha_s^2\int d\xi\, n(\xi)$,
thereby reproducing the well-known nuclear shadowing result of
Nikolaev and Zakharov. For $N=1$, we have to expand all Green's
functions in (\ref{4.1}) and then
to collect all contributions to order $O(A_0^2)$.

\begin{figure}[h]\epsfxsize=7.7cm 
\centerline{\epsfbox{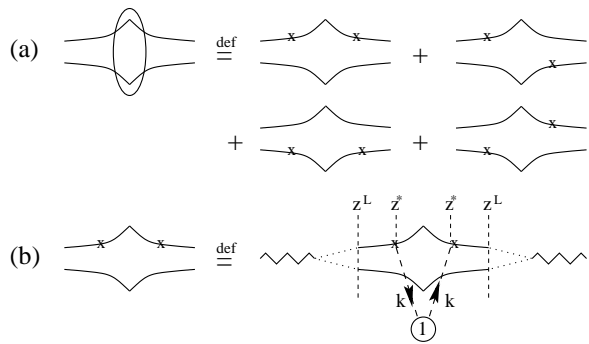}}
\vspace{0.5cm}
\caption{(a) Diagrammatic representation of the $N=1$ contribution
to the total photoabsorption cross section (\protect\ref{4.1}). 
For each term, the upper (lower) line denotes the Green's function 
with energy $p_1$ ($p_2$) and the cusps denote the position of the 
cut. (b) Definition of the diagrammatic shorthand used in (a) and
calculated in (\protect\ref{4.5}). Crosses stand for one power of 
the scattering potential $A_0$. The medium average ensures that the
momentum flow $k$ through the cut is conserved and that the potential
$A_0$ is linked to the $q$-$\bar{q}$-system in $M_{\rm fi}$ and
$M_{\rm fi}^\dagger$ at the same longitudinal position.
}\label{fig4}
\end{figure}

The zeroth and first order contributions $O(A_0^0)$ and $O(A_0^1)$ to 
$\sigma_{total}^{\gamma^*\to q\, \bar{q}}$ vanishes due to 
energy-momentum conservation: without momentum transfer to
the medium, the $q$-$\bar{q}$-pair cannot appear on-shell.
To second order $O(A_0^2)$,
there are four different terms which we depict diagrammatically
in Fig.~\ref{fig4}a: for each term, the full lines  
correspond to the full lines shown in Fig.~\ref{fig3}
and describe the full Green's functions in the total photoabsorption 
cross section (\ref{4.1}). To simplify the representation,
we have dropped in comparison to  Fig.~\ref{fig3} the photon 
lines and incoming $q$-$\bar{q}$ wavefunctions. 
For the total photoabsorption cross section, the transverse positions 
at the cut are equal for the amplitude and the complex conjugate amplitude.
%
%
Let us consider the first term on the r.h.s. of 
Fig.~\ref{fig4}. Here, the Green's functions of argument $p_2$
are free while both Green's functions with argument $p_1$ are
expanded to first order. Averaging over the position 
$(\check{\bf r}_1\, ,\check{z}_1)$ of the center of the 
scattering potential, we find
\begin{eqnarray}
  &&\Big\langle \int 
  {\it d}{\bf x}_\perp\, \bar G_0({\bf \bar b}_2;{\bf x}\vert p_2)\,
      \bar G_0({\bf x};{\bf b}_2\vert p_2)
    \nonumber \\
 && \qquad \quad \times \int 
      {\it d}{\bf x'}_\perp\, 
    \bar G^{(1)}({\bf b}_1;{\bf x'}\vert p_1)\,
    \bar G^{(1)}({\bf x'};{\bf \bar b}_1\vert p_1)
    \Big\rangle
    \nonumber \\
 && = \frac{1}{A_\perp} \int d\xi_1\, n(\xi_1)\, 
      \int \frac{d\bbox{\kappa}_\perp}{(2\pi)^2}\, 
      \vert a_0(\bbox{\kappa}_\perp)\vert^2\,
      T_{a_1}\, T_{a_1} 
    \nonumber \\
 && \qquad \times \delta^{(2)}({\bf \bar b}_2-{\bf b}_2)\, 
      \delta^{(2)}({\bf \bar b}_1-{\bf b}_1)\, .
  \label{4.5}
\end{eqnarray}   
This term does not depend on the separation between the $q$-$\bar{q}$
pair since the potential touches only one of the quark lines. In
contrast, the second term on the r.h.s. of Fig.~\ref{fig4}a depends
on this separation and takes the explicit form
\begin{eqnarray}
 && - \int d\xi_1\, n(\xi_1)\, 
      \int \frac{d\bbox{\kappa}_\perp}{(2\pi)^2}\,
      \vert a_0(\bbox{\kappa}_\perp)\vert^2\, T_{a_1}\, T_{a_1}\, 
      e^{-i\bbox{\kappa}_\perp\cdot ({\bf b}_1-{\bf b}_2)}
    \nonumber \\
 && \qquad \times 
      e^{-i\frac{\bbox{\kappa}_\perp^2}{2\nu\alpha(1-\alpha)}(\xi_1-z_L)}\, 
      \delta^{(2)}\left({\bf \bar b}_2-{\bf b}_2 - 
     \frac{\bbox{\kappa}_\perp}{p_2}(\xi_1-z_L)\right)
    \nonumber \\
 && \qquad \times 
      \delta^{(2)}\left({\bf \bar b}_1-{\bf b}_1 - 
     \frac{\bbox{\kappa}_\perp}{p_1}(\xi_1-z_L) \right) 
     \, \frac{1}{A_\perp}\, .
  \label{4.6}
\end{eqnarray}   
This shows explicitly that the transverse Brownian motion induced
by rescattering is taken into account in our formulation: the
$q$-$\bar{q}$ dipole does not propagate at fixed transverse
separation ${\bf \bar b}_1-{\bf b}_1$ but changes its size
in response to an interaction at the longitudinal position $\xi_1$
by terms of the form $\frac{\bbox{\kappa}_\perp}{p_1}(\xi_1-z_L)$.
For $N=1$, the first and only interaction takes place by definition
at the position $z_L = \xi_1$ and these terms vanish. For $N>1$,
however, they do not vanish and lead to a nontrivial dynamical
evolution of the transverse dipole size, see below.

 Adding up the four real contributions in Fig.~\ref{fig4}a,
and doing the colour trace, we find 
\begin{eqnarray}
  \sigma_{total}^{\gamma^*\to q\, \bar{q}}(N=1)
  &=& N_c\, \alpha_{\rm em}\, \int d\xi\, n(\xi)\, \int d\alpha\, 
  \nonumber \\
  && \times \int {\it d}{\bf b}\, 
    \Phi({\bf b}; {\bf b};\alpha)\,\, 
    \sigma({\bf b})\, ,
    \label{4.7} 
\end{eqnarray}
\begin{equation}
 \sigma({\bf b})  = 2\, C_F
      \int \frac{d\bbox{\kappa}_\perp}{(2\pi)^2}\, 
      \vert a_0(\bbox{\kappa}_\perp)\vert^2\,
      \left(1- e^{-i\bbox{\kappa}_\perp\cdot{\bf b} }\right)\, .
  \label{4.8}
\end{equation}   
This is the result of Nikolaev and Zakharov~\cite{NZ91}. 
The total cross section (\ref{4.1}) is the product of the Born probability 
$ \Phi({\bf b}; {\bf b};\alpha)$ for the incoming hadronic
Fock state times a dipole cross section $\sigma({\bf b})$,
integrated over the transverse size ${\bf b}$ of the dipole
and the energy distribution $\alpha$ between the quark and antiquark.

\subsection{$N=2$ tagged: non-trivial colour interference effects} 
\label{sec4b}
\begin{figure}[h]\epsfxsize=8.7cm 
\centerline{\epsfbox{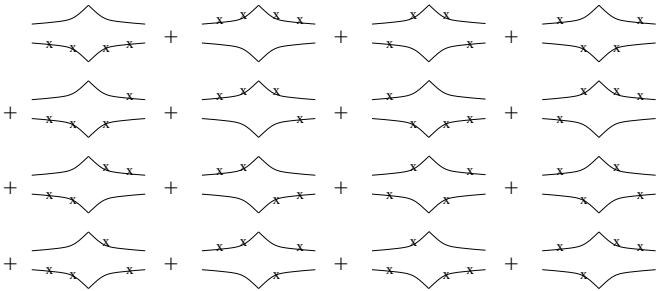}}
\vspace{0.5cm}
\caption{The 16 contributions to the $N=2$ photoabsorption
cross section (\protect\ref{4.9}). In this ``tagged'' case,
exactly two real momentum transfers occur on the amplitude
level. The tagging gives rise to non-trivial colour 
interference effects.
}\label{fig5}
\end{figure}

For more than $N=1$ scattering center, there are in general non-vanishing
interference terms between amplitudes containing different powers of
$A_0$, see section~\ref{sec4c} below. Such contributions appear in
the generic situation in which one knows the distribution of scattering
centers but no information is available on whether these centers have
participated with finite momentum transfer in the scattering process.
At least in a gedanken experiment, however, this additional information
can be obtained from the recoil of each scattering center, i.e., by
{\it tagging} the scattering centers. For this very exclusive observable,
all contributions to the cross section contain the same power of $A_0$
in both amplitudes, and all scattering centers contribute to the amplitude
with finite momentum transfers. We say, such scattering centers contribute
with {\it real terms}, in contrast to the {\it contact term} contributions
which we discuss in section~\ref{sec4c}. Tagged scenarios were considered
recently in the study of the non-abelian LPM-effect~\cite{GLV99}, and
there is some interest in comparing results of tagged and untagged
scenarios to understand the origin of colour triviality. To this aim,
we consider here as a second application of the opacity expansion 
the case of two tagged scattering centers: (i) two scattering 
centers are involved in the process (\ref{4.1}) and (ii) both scattering
centers provide {\it real terms}, i.e., transfer finite transverse
momentum on the amplitude level. The 16 contributions to this
${\sigma_{\rm tagged}^{\gamma^*\to q\, \bar{q}}}(N=2)$
are listed in Fig.~\ref{fig5}. Summing them up, we find
\begin{eqnarray}
  &&{\sigma_{\rm tagged}^{\gamma^*\to q\, \bar{q}}}(N=2)
  \nonumber \\
  \qquad &=& \alpha_{\rm em} 
     \int d\xi_1\, n(\xi_1)\, \int_{\xi_1}  d\xi_2\, n(\xi_2)\, 
     \int d\alpha \int d{\bf b}
  \nonumber \\
  \qquad && \times   \int \frac{d\bbox{\kappa}_{1,\perp}}{(2\pi)^2}\, 
  \frac{d\bbox{\kappa}_{2,\perp}}{(2\pi)^2}\, 
      \vert a_0(\bbox{\kappa}_{1,\perp})\vert^2\,
      \vert a_0(\bbox{\kappa}_{2,\perp})\vert^2
   \nonumber \\
  \qquad && \times \left({\rm Tr}\lbrack a\, a\, b\, b\, \rbrack \, 
       A^{(1)} + {\rm Tr}\lbrack a\, b\, a\, b\, \rbrack \,
        A^{(2)}\right)\, ,
      \label{4.9}
  \end{eqnarray}
where we have introduced the notational shorthands
\begin{eqnarray}
  A^{(1)} &=& 
    4\, \Phi({\bf b},{\bf b};\alpha)\, 
    \left(1 - e^{-i\,{\bbox{\kappa}}_{1,\perp}\cdot{\bf b}}\right)\, ,
   \label{4.10} \\
  A^{(2)} &=& 4\, 
  \Phi\left({\bf b}-\frac{\bbox{\kappa}_{2,\perp}(\xi_2-\xi_1)}
  {2\, \nu\, \alpha\, (1-\alpha)},
  {\bf b}+\frac{\bbox{\kappa}_{2,\perp}(\xi_2-\xi_1)}
  {2\, \nu\, \alpha\, (1-\alpha)};\alpha\right)
  \nonumber \\
  && \times \left( 
      e^{-i\,{\bbox{\kappa}}_{1,\perp}\cdot{\bf b}}
      e^{-i\,\frac{{\bbox{\kappa}}_{1,\perp}\cdot {\bbox{\kappa}}_{2,\perp}}
         {2\nu\alpha(1-\alpha)} (1-2\alpha)\, (\xi_2-\xi_1)} 
        \right.
  \nonumber \\
  && \qquad \left. 
  - e^{-i\,{\bbox{\kappa}}_{1,\perp}\cdot {\bbox{\kappa}}_{2,\perp}
      \frac{(\xi_2-\xi_1)}{\nu\alpha}} \right)\,
      e^{-i\,{\bbox{\kappa}}_{2,\perp}\cdot{\bf b}}\, .
   \label{4.11} 
\end{eqnarray}  
Here, the term $A^{(2)}$ contains information about the non-trivial
dynamical evolution of the $q$-$\bar{q}$-separation which is 
determined by the ``relative transverse mass''
$\mu = \left(\frac{1}{p_1} - \frac{1}{p_2}\right)^{-1} = 
\nu\alpha(1-\alpha)$. The momentum transfer ${\bbox{\kappa}}_{2,\perp}$
determines this change of the $q$-$\bar{q}$-size via a classical
equation of motion: $\propto \frac{{\bbox{\kappa}}_{2,\perp}}{\mu}
\left(\xi_2-\xi_1\right)$. In the high energy limit $\nu\to\infty$,
this transverse motion can be neglected and the $q$-$\bar{q}$-pair
propagates at fixed separation. We find
\begin{eqnarray}
  \lim_{\nu\to\infty} 
  &&\sigma_{\rm tagged}^{\gamma^*\to q\, \bar{q}}(N=2)
  \nonumber \\
  &=& \alpha_{\rm em}\,  
     \int d\xi_1\, n(\xi_1)\, \int_{\xi_1}  d\xi_2\, n(\xi_2)\, 
     \int d\alpha \int d{\bf b}
  \nonumber \\
  && \times   \int \frac{d\bbox{\kappa}_{1,\perp}}{(2\pi)^2}\, 
  \frac{d\bbox{\kappa}_{2,\perp}}{(2\pi)^2}\, 
      \vert a_0(\bbox{\kappa}_{1,\perp})\vert^2\,
      \vert a_0(\bbox{\kappa}_{2,\perp})\vert^2
   \nonumber \\
  && \times \Phi({\bf b},{\bf b};\alpha)\, 4\, 
  \left(1 - e^{-i\,{\bbox{\kappa}}_{1,\perp}\cdot{\bf b}}\right)
  \nonumber \\
  && \times \left({\rm Tr}\lbrack a\, a\, b\, b\, \rbrack -
      {\rm Tr}\lbrack a\, b\, a\, b\, \rbrack \,
        e^{-i\,{\bbox{\kappa}}_{2,\perp}\cdot{\bf b}}\right)\, .
      \label{4.12}
  \end{eqnarray}
Even in this limiting case, 
the colour algebra does not factorize from the momentum
dependence. This is a consequence of the non-trivial colour 
interference between different photodissociation amplitudes.
The photoabsorption cross section cannot be
written as a function of the dipole cross section 
$\sigma({b})$. We note that this complication stems entirely
from the non-abelianess of the problem. If we replace both
colour traces in (\ref{4.12}) by the same constant, 
the two ${\bbox{\kappa}}_{i,\perp}$-integrations combine to the
square of the dipole cross section (\ref{4.8}). To sum up:
in the tagged case, non-trivial colour interference effects
prevent us from representing the photoabsorption cross section
as a function of the dipole cross section.

%
\subsection{Contact terms}
\label{sec4c}

\begin{figure}[h]\epsfxsize=8.7cm 
\centerline{\epsfbox{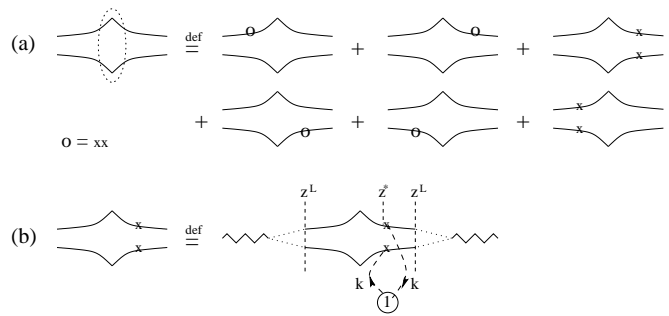}}
\vspace{0.5cm}
\caption{
(a) The six contact terms for the $N=1$ scattering cross section.
(b) Definition of the diagrammatic shorthand used in (a). The medium
average ensures that (i) the momentum flow $k$ is conserved and no 
net momentum is transferred to the $q$-$\bar{q}$-system, and that 
(ii) both powers of the scattering potential act at the same
longitudinal position $z^*$.
}\label{fig6}
\end{figure}

In the calculation of the $N=1$ cross section (\ref{4.7}), we did
not include all terms of the Taylor expansion of (\ref{4.1}) to
order $O(A_0^2)$. A second order contribution which was not 
included is e.g.
\begin{eqnarray}
  &&\Big\langle \int 
  {\it d}{\bf x}_\perp\, \bar G_0({\bf \bar b}_2;{\bf x}\vert p_2)\,
      \bar G_0({\bf x};{\bf b}_2\vert p_2)
    \nonumber \\
 && \qquad \quad \times \int 
      {\it d}{\bf x'}_\perp\, 
    \bar G_0({\bf b}_1;{\bf x'}\vert p_1)\,
    \bar G^{(2)}({\bf x'};{\bf \bar b}_1\vert p_1)
    \Big\rangle
    \nonumber \\
 && = - \frac{1}{2}\, 
      \int d\xi_1\, n(\xi_1)\,
      \int \frac{d\bbox{\kappa}_\perp}{(2\pi)^2}\, 
      \vert a_0(\bbox{\kappa}_\perp)\vert^2\,
      T_{a_1}\, T_{a_1} 
    \nonumber \\
 && \qquad \times \delta^{(2)}({\bf \bar b}_2-{\bf b}_2)\, 
      \delta^{(2)}({\bf \bar b}_1-{\bf b}_1)\, .
  \label{4.13}
\end{eqnarray}   
Here, the potential $A_0$ is linked two times to the complex 
conjugate amplitude and zero times to the amplitude. The medium
average $\langle \dots\rangle$ results in two
important constraints: (i) both powers of $A_0$ couple to the 
$q$-$\bar{q}$-pair at the same longitudinal position $\xi_1$,
and (ii) no net momentum is transferred to the $q$-$\bar{q}$-
system. Scattering contributions with these properties were
called {\it virtual} by Mueller and collaborators~\cite{M90,BDMPS1}. 
We refer to them as {\it contact terms}, all other interactions are
refered to as {\it real}. Diagrammatically, the
six contact terms for the $N=1$ case are given in Fig.~\ref{fig6}a,
and equation (\ref{4.13}) is represented by the first diagram
on the r.h.s.

%
For the $N=1$ photoabsorption cross section (\ref{4.7}), contact 
terms do not contribute since at least one interaction with the
$q$-$\bar{q}$-system is needed in both the amplitude and complex
conjugate amplitude in order to get the final state on shell. 
For $N>2$, the same argument about energy-momentum conservation
does not imply the absence of contact terms. We have to 
distinguish the following cases:
\begin{enumerate}
\item 
  $\sigma_{\rm inel.}^{\gamma^*\to q\bar{q}}$ inelastic:
  the cross section involves at least one real interaction.
\item
  $\sigma_{\rm diff.}^{\gamma^*\to q\bar{q}}$ diffractive (or elastic):
  the medium interacts with the $q$-$\bar{q}$-pair only
  via contact terms.
\item
  $\sigma_{\rm total}^{\gamma^*\to q\bar{q}}$ total:
  the cross section involves at least one (real or contact)
  interaction in both $M_{fi}$ and $M_{fi}^\dagger$.
\end{enumerate}
The notion "inelastic" is justified since any real interaction
changes the colour of the target and thus affects the hadronic
activity between projectile and target rapidity (in this sense,
the target "breaks up"). The notion ``diffractive'' or ``elastic''
involves an additional
assumption: by construction, contact terms transfer exactly zero
net transverse momentum to the $q$-$\bar{q}$-pair. Also, we 
regard in all calculations and for arbitrary scatterings the longitudinal  
momentum transfer as sufficiently small to justify the approximation
$a_0(\bbox{\kappa}) \approx a_0(\bbox{\kappa}_\perp)$.
Our assumption in the above definitions of the diffractive and
total photoabsorption cross sections is that despite neglecting
the longitudinal momentum transfer in all explicit calculations,
it is sufficient to put the (extremely weakly off-shell) partonic
contributions of $\gamma^*$ on-shell. This is usually assumed in
the calculation of the total~\cite{M90} and diffractive~\cite{KL99} 
cross sections.

For $N>1$, contact terms obviously play an important role in the
calculation of the inelastic, diffractive and total photoabsorption
cross section.  In the remainder of this subsection, we 
summarize some of their important properties and shorthands, 
which will be heavily used in the following subsections:

\subsubsection{Identities involving contact terms} 

\begin{figure}[h]\epsfxsize=7.7cm 
\centerline{\epsfbox{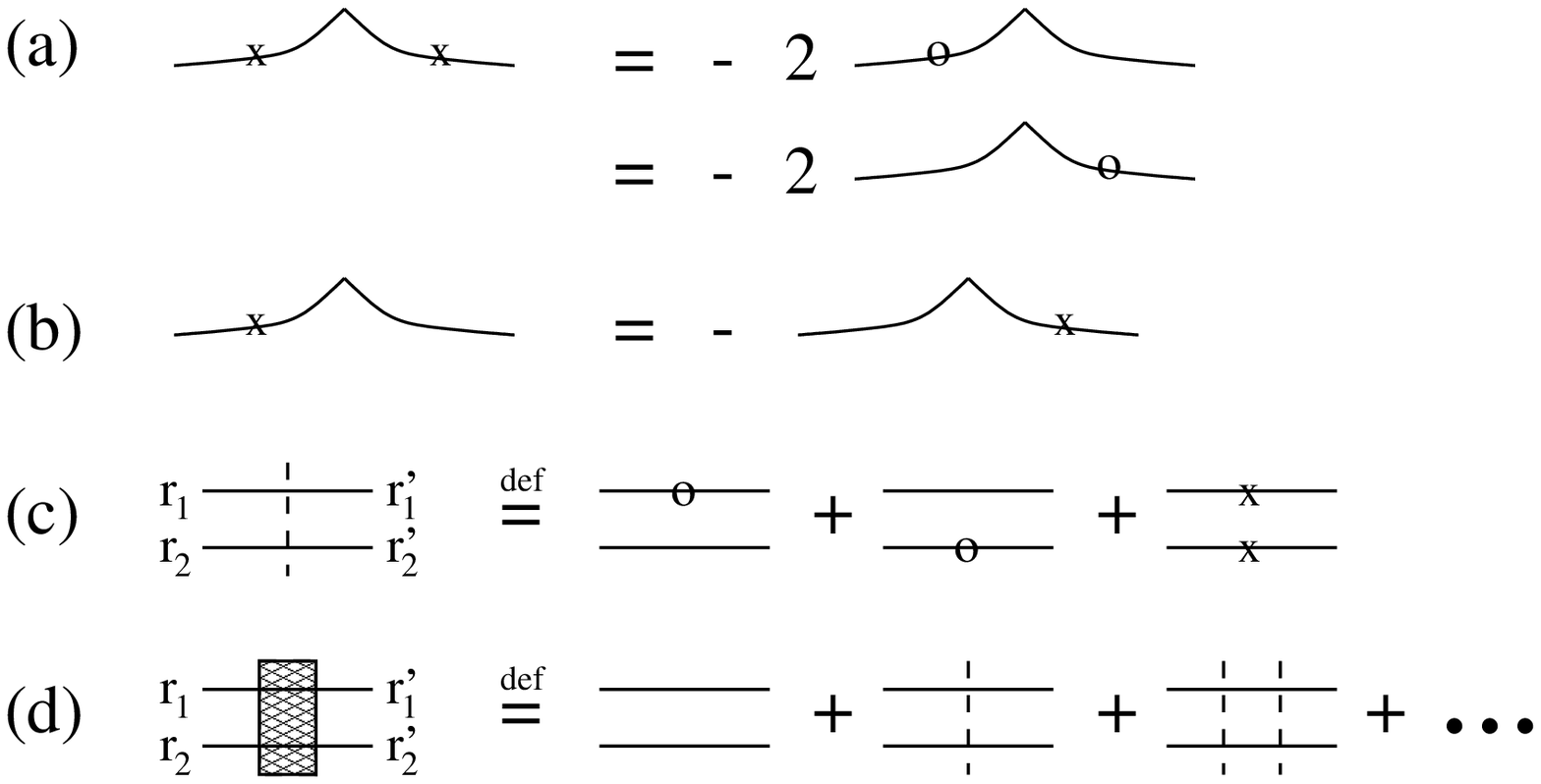}}
\vspace{0.5cm}
\caption{At the cut, there are simple identities between real
interaction terms and contact terms (a), and amongst real
interaction terms (b). (c) A combination
of the three possible contact terms at the same longitudinal 
position $\xi$ leads to a dipole cross section $\sigma({\bf r}_1(\xi)
-{\bf r}_2(\xi))$ whose transverse size is evolved further with free 
Greens functions, see (\protect\ref{4.14})
(d) Iterating the combination of contact terms (c) leads to a 
dynamical evolution of the dipole size between initial transverse
size ${\bf r}_1-{\bf r}_2$ and final size ${\bf r}_1'-{\bf r}_2'$.
This evolution is described by the path-integral 
(\protect\ref{4.15}).
}\label{fig7}
\end{figure}

First we observe that in (\ref{4.13}), the Green's functions of 
momentum $p_2$ result in the
transverse $\delta$-function $\delta^{(2)}({\bf \bar b}_2-{\bf b}_2)$.
Dropping them on both sides of (\ref{4.13}), and observing that
(\ref{4.13}) and (\ref{4.5}) differ by a factor $\frac{-1}{2}$,
we have checked the identity in Fig.~\ref{fig7}a. By a
similar calculation, one also checks the identity Fig.~\ref{fig7}b.
%
%
\subsubsection{A path integral from iterating contact terms} 

We now consider the case in Fig.~\ref{fig7}c, where the 
$q$-$\bar{q}$-state is evolved with exactly one contact term
between longitudinal positions
$z$ and $z'$ from a transverse separation ${\bf r}_1-{\bf r}_2$
to a transverse separation ${\bf r}_1'-{\bf r}_2'$. The 
sum over the three possible contact terms at longitudinal position
$\xi$, represented on the r.h.s. of  Fig.~\ref{fig7}c,
takes the explicit form
\begin{eqnarray}
  && - \frac{T_a\,\cdots T_a}{2\, C_f}\int d\xi\, n(\xi)\, 
  \int d{\bf r}_1(\xi)\, d{\bf r}_2(\xi)\, 
  \sigma\left({\bf r}_1(\xi)-{\bf r}_2(\xi)\right)
  \nonumber \\
  && \times\, \bar G_0({\bf r}_2,z;{\bf r}_2(\xi),\xi|p_2)\, 
  \bar G_0({\bf r}_2(\xi),\xi;{\bf r}_2',z'|p_2)
  \nonumber \\
  && \times\, \bar G_0({\bf r}_1',z';{\bf r}_1(\xi),\xi|p_1)\, 
  \bar G_0({\bf r}_1(\xi),\xi;{\bf r}_2,z|p_1)\, . 
  \label{4.14}
\end{eqnarray}  
Since Fig.~\ref{fig7}c is only part of a more complicated
Feynman diagram, we can say nothing about the positioning of the
colour factors $T_a$. The corresponding expression $T_a\,\cdots T_a$
in (\ref{4.14}) is purely formal. However, for the following, we 
anticipate an argument explained in the next subsection: for our 
purposes, the colour factors reduce to a Casimir factor $C_F$, i.e., 
we can substitute in (\ref{4.14}) $\frac{T_a\,\cdots T_a}{C_f} \to 1$. 
Then, equation (\ref{4.14}) is seen to correspond to the first order
$n(\xi)$ density expansion of the double path integral (for $z'>z$)
\begin{eqnarray}
  && M({\bf r}_1,{\bf r}_2,z;{\bf r}_1',{\bf r}_2',z'|\alpha|\nu)
  \nonumber \\
  && \qquad = 
  \int\limits_{{\bf r}_1(z)={\bf r}_1}^{{\bf r}_1(z')={\bf r_1'}} 
  {\cal D}{\bf r}_1\, 
  \int\limits_{{\bf r}_2(z)={\bf r}_2}^{{\bf r}_2(z')={\bf r_2'}}
  {\cal D}{\bf r}_2
  \nonumber \\
  && \qquad \quad \times
     \exp\Big\lbrace i\int_{z}^{z'} d\bar{\xi}\, \frac{\nu}{2}
     \left( \alpha\, \dot {\bf r}_2^2 + (1-\alpha)\, \dot {\bf r}_1^2
       \right) 
  \nonumber \\
  && \qquad \qquad \qquad \qquad + i\, \frac{1}{2}\, n(\xi)\, 
       \sigma\left({\bf r}_1(\xi) - {\bf r}_2(\xi)\right)
       \Big\rbrace \, .
  \label{4.15}
\end{eqnarray}
Moreover, the expansion of (\ref{4.15}) to $m$-th order in the density
$n(\xi)$ corresponds exactly to the $m$-th term on the r.h.s. of 
Fig.~\ref{fig7}d. This shows that the iteration
Fig.~\ref{fig7}d of the one contact term interaction Fig.~\ref{fig7}d
is described by the path-integral (\ref{4.15}). Since this path integral
plays an important role in what follows, we simplify it further.
Using the coordinate transformation
\begin{eqnarray}
  {\bf r}_a(\xi) &=& (1-\alpha)\, {\bf r}_1(\xi) + \alpha\, {\bf r}_2(\xi)\, ,
  \label{4.16}\\
  {\bf r}_b(\xi) &=& {\bf r}_1(\xi) - {\bf r}_2(\xi)\, ,
  \label{4.17}
\end{eqnarray}
we can write
\begin{eqnarray}
  && M({\bf r}_1,{\bf r}_2,z;{\bf r}_1',{\bf r}_2',z'|\alpha|\nu)
  \nonumber \\
  && = {\cal K}_0\bigl({\bf r}_a(z'),z';{\bf r}_a(z),z|\nu\bigr)
  \nonumber \\
  && \quad \times {\cal K}\bigl({\bf r}_b(z'),z';{\bf r}_b(z),z
       |\nu\alpha(1-\alpha)\bigr)\, .
  \label{4.18}
\end{eqnarray}
The path-integral on the r.h.s. of this expression is given by
(for $z'>z$)
\begin{eqnarray}
  &&{\cal K}\bigl({\bf r}(z'),z';{\bf r}(z),z|\mu\bigr) 
  \nonumber \\
  && \quad =   \int {\cal D}{\bf r}\, 
  \exp\left\{i\, \int\limits_z^{z'}\, d\xi\,
  \left[{\mu\over 2}\dot{\bf r}^2 
  + i\, \frac{1}{2} n(\xi)\, \sigma\left({\bf r}\right)
  \right] \right\}\, .
  \label{4.19}
\end{eqnarray}
Its opacity expansion reads 
\begin{eqnarray}
 &&{\cal K}({\bf r}',z';{\bf r},z) =
    {\cal K}_0({\bf r}',z';{\bf r},z)
 \nonumber \\
 && - \int\limits_{z}^{z'}\, {\it d}\xi \int {\it d}\bbox{\rho}\,
 {\cal K}_0({\bf r}',z';{\bf r},\xi)\,
   \Sigma\bigl(\bbox{\rho},\xi\bigr)\, 
   {\cal K}_0(\bbox{\rho},\xi;{\bf r},z) 
 \nonumber \\
 && + \int\limits_{z}^{z'} {\it d}\xi_1\,
    \int\limits_{\xi_1}^{z'} {\it d}\xi_2\,
    \int {\it d}\bbox{\rho}_1\,{\it d}\bbox{\rho}_2\,
    {\cal K}_0({\bf r}',z';\bbox{\rho}_2,\xi_2)\, 
    \Sigma\bigl(\bbox{\rho}_2,\xi_2\bigr)
    \nonumber \\
 && \qquad \times 
    {\cal K}(\bbox{\rho}_2,\xi_2;\bbox{\rho}_1,\xi_1)\, 
    \Sigma\bigl(\bbox{\rho}_1,\xi_1 \bigr)\,
    {\cal K}_0(\bbox{\rho}_1,\xi_1;{\bf r},z)\, , 
 \label{4.20} 
\end{eqnarray}
where $\Sigma\bigl(\bbox{\rho},\xi\bigr)= \frac{1}{2}\, n(\xi)
\sigma\bigl(\bbox{\rho}\bigr)$. Here, we have suppressed the explicit 
$\mu$-dependence in ${\cal K}$. The corresponding free Green's 
function ${\cal K}_0$ reads
\begin{equation}
  {\cal K}_0({\bf r}',z';{\bf r},z)
  = \frac{\mu}{2\pi\, i\, (z'-z)}
    \exp\left\{ { {i\mu}
           \left({\bf r}' - {\bf r}\right)^2 \over {2\, (z'-z)} }
           \right\}\, .
  \label{4.21}
\end{equation}
In analogy to our definition
of the Green's function $\bar G$ in (\ref{2.16}),
we define ${\cal K}\bigl({\bf r}(z),z;{\bf r}(z'),z'|\mu\bigr)
\equiv {\cal K}^\dagger \bigl({\bf r}(z'),z';{\bf r}(z),z|\mu\bigr)$
for $z'>z$. 

A path integral ${\cal K}$ of the form derived here was first used
by Zakharov in the abelian problem of calculating the passage of
ultrarelativistic positronium through matter~\cite{Z87}. More
recently, the same path integral was shown to appear in the
final expression of the medium-dependence of the LPM-bremsstrahlung
spectrum~\cite{Z96,WG99}.

\subsection{$N=$ arbitrary, untagged: contact terms remove colour 
interference effects}
\label{sec4d}

\begin{figure}[h]\epsfxsize=8.7cm 
\centerline{\epsfbox{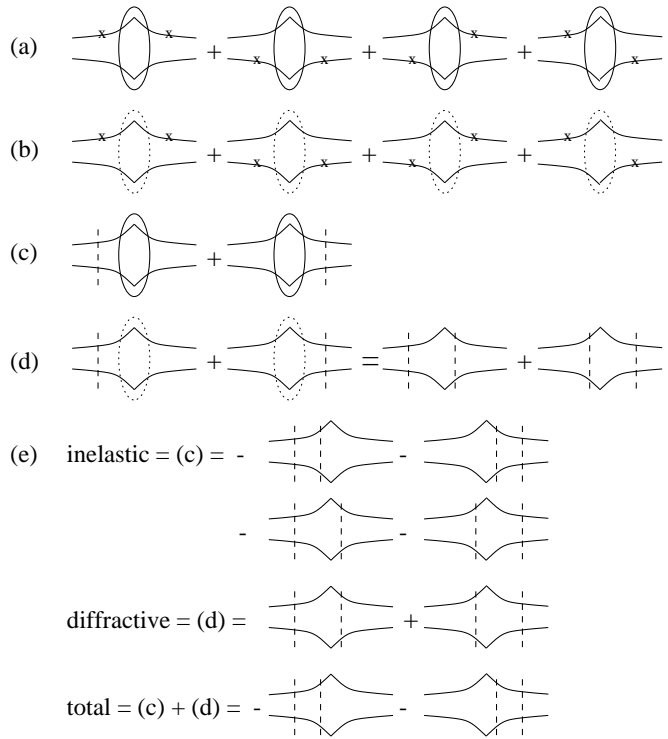}}
\vspace{0.5cm}
\caption{The contributions to the $N=2$ photoabsorption
cross section: (a) the 16 real contributions
which are present in the ``tagged'' case. (b) Contributions for
which the first interaction is real and the second a contact term.
(c) Contributions for which the first interaction is a contact 
term and the second interaction transfers momentum. (d) Terms
which contain two contact interactions. Here, energy-momentum
conservation ensures that the two contact terms do not stand on 
the same side of the cut. (e) Inelastic, diffractive and total 
photoabsorption cross section are determined by different 
combinations of the contributions (c) and (d). Rewriting the
real interaction in (c) as the sum of two negative contact term
contributions, a particularly simple representation is obtained. 
}\label{fig8}
\end{figure}

What happens if contact terms are included in the $N=2$ calculation
of the elastic, inelastic and total photoabsorption cross section ?
In Fig.~\ref{fig8}, we have classified into four subsets all 
terms of order $O(n(\xi)^2)$ which have to be considered for
the $N=2$ photoabsorption cross sections. There are in 
Fig.~\ref{fig8}a the $4\times 4 = 16$ terms
for which both interactions are real, in Fig.~\ref{fig8}b the
$4\times 6=24$ terms for which only the second interaction is
contact, in Fig.~\ref{fig8}c the $3\times 4\times 2$ terms
for which only the first interaction is contact, and in
Fig.~\ref{fig8}d the $2\times 3\times 6 = 36$ which
involve two contact interactions. All together, these
are 100 terms. For the total and the inelastic cross section,
the terms in Fig.~\ref{fig8}a and Fig.~\ref{fig8}b both
contribute, but they cancel each other exactly. More precisely, 
the first diagram in Fig.~\ref{fig8}a cancels the first diagram 
in Fig.~\ref{fig8}b, the second diagram in Fig.~\ref{fig8}a 
cancels the second diagram in Fig.~\ref{fig8}b, etc. This is
a consequence of the identities Fig.~\ref{fig7}a and ~\ref{fig7}b
which we apply to the last interaction before the cut.

\begin{figure}[h]\epsfxsize=8.7cm 
\centerline{\epsfbox{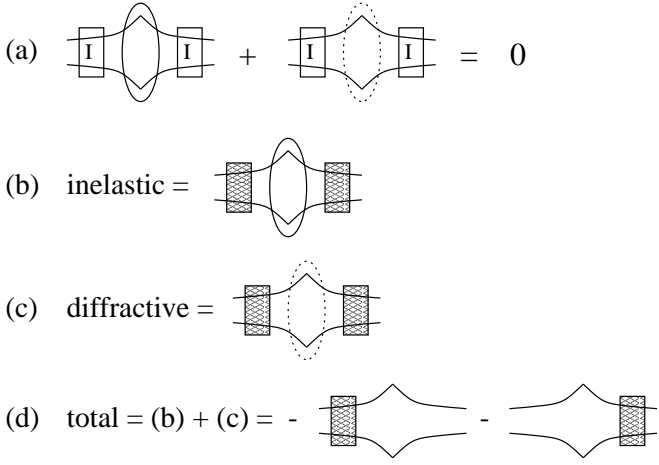}}
\vspace{0.5cm}
\caption{(a) Cancellation of real and contact terms for any
combination of interactions [I] which contains at least one
real or contact interaction. (b)-(d) The identity (a) allows for
the generalization to arbitrary $N$ of the $N=2$ photoabsorption
cross section of Fig.~\protect\ref{fig8}.
}\label{fig9}
\end{figure}

The same argument cannot be made for the cancellation of the
diagrams in Fig.~\ref{fig8}c and Fig.~\ref{fig8}d:
energy momentum conservation restricts the occurence of double
contact terms, since at least one interaction is needed on the
amplitude level to get the final state on-shell.
Depending on whether we calculate the diffractive, inelastic or
total photoabsorption cross section, different combinations
of these two sets of diagrams contribute. 
The real interaction Fig.~\ref{fig4}a at the cut is determined
by minus the combination Fig.~\ref{fig6}a of contact terms. This
allows us to express all three contributions to the photoabsorption 
cross section in terms of contact terms, as shown in Fig.~\ref{fig8}. 
Most importantly, 
all contributions to these photoabsorption cross sections 
have the colour trace ${\rm Tr}[aabb]$: for the untagged case for
which contact terms are included, colour interference terms 
vanish in the diffractive,
inelastic and total $N=2$ photoabsorption cross section.
%
The above argument can be generalized easily to arbitrary $N$.
As shown in Fig.~\ref{fig9}a, whenever at least one of the
$(N-1)$ first interactions is real, the sum of all real and 
virtual terms for the $N$-th interaction cancels. Inelastic,
diffractive and total photoabsorption cross section can thus
again be described in terms of $(N-1)$ contact terms and
a characteristic $N$-th contribution at the cut. As an
immediate consequence, the colour trace of all these contributions
reads
\begin{equation}
  {\rm Tr}\lbrack T_{a_1}\, T_{a_2}\, \dots T_{a_N}\, T_{a_N}\, 
                  T_{a_{N-1}}\,\dots T_{a_1}\rbrack = N_c\, C_F^N\, .
  \label{4.22}
\end{equation}
In contrast to the tagged photoabsorption cross section, the 
colour structure factorizes from the momentum dependence:
Inclusion of the contact terms results in colour triviality
for the $N$-fold diffractive, inelastic and total photoabsorption
cross section. This justifies the assumption made in deriving 
the path integral (\ref{4.15}) in the last subsection.

%

According to Fig.~\ref{fig9}(b)-(d), the inelastic, diffractive and
total photoabsorption cross section can be written in terms of the 
path-integral ${\cal K}$ of (\ref{4.19}) which describes the 
iteration of contact terms. For the inelastic photoabsorption 
cross section, this leads to the expression
\begin{eqnarray}
  \sigma_{\rm inel.}^{\gamma^*\to q\, \bar{q}}
  &=& N_c\, \alpha_{\rm em}\, \int d\alpha\, 
    \int {\it d}{\bf b}\, {\it d}{\bf \bar b}
    \nonumber \\
 && \times 
    \Phi\left({\bf b};{\bf \bar b};\alpha\right)
      \int_{\xi_{nl}}^L {\it d}\xi\, n(\xi)\, \int {\it d}{\bf r}(\xi)
    \nonumber \\
 && \times 
      {\cal K}\bigl({\bf \bar b},0;{\bf r}(\xi),\xi
       |\nu\alpha(1-\alpha)\bigr)\, 
       \sigma\left({\bf r}(\xi)\right)\, 
    \nonumber \\
 && \times 
    {\cal K}\bigl({\bf r}(\xi),\xi;{\bf b},0
       |\nu\alpha(1-\alpha)\bigr)\, .
  \label{4.23}
\end{eqnarray}   
%
%
%
Here, the Green's functions ${\cal K}$ describe the propagation of
the $q$-$\bar{q}$-dipole from the front end of the target at $z=0$
up to the position $\xi$ of the last interaction. The lower boundary
$\xi_{nl}$ of the integration over $\xi$ is defined by the position
of the next to last (``nl'') interaction. This is an awkward  
property, since we have to expand the ${\cal K}$'s in the number $N$
of interactions to determine  $\xi_{nl}$ order for order in $N$.
This makes it important to find for (\ref{4.23}) with the 
help of the opacity expansion (\ref{4.20}) the equivalent representation
\begin{eqnarray}
  &&\sigma_{\rm inel.}^{\gamma^*\to q\, \bar{q}}
  = N_c\, \alpha_{\rm em}\, \int d\alpha\, 
    \int {\it d}{\bf b}\, {\it d}{\bf \bar b}\, 
    \Phi\left({\bf b};{\bf \bar b};\alpha\right)\, 
  \nonumber \\
  &&\qquad \times
    \left[ \delta^{(2)}\left({\bf b}- {\bf \bar b}\right) \right.
    \nonumber \\
 &&\qquad \quad \left.
     - \int d{\bf r}_e\, 
      {\cal K}\bigl({\bf \bar b},0;{\bf r}_e,L|\mu\bigr)\, 
    {\cal K}\bigl({\bf r}_e,L;{\bf b},0|\mu\bigr) \right]\, .
  \label{4.24}
\end{eqnarray}   
For the diffractive contribution, one finds
\begin{eqnarray}
  &&\sigma_{\rm diff.}^{\gamma^*\to q\, \bar{q}}
  = N_c\, \alpha_{\rm em}\, \int d\alpha\, 
    \int {\it d}{\bf b}\, {\it d}{\bf \bar b}\, 
    \Phi\left({\bf b};{\bf \bar b};\alpha\right)\, 
    \int d{\bf r}_e
    \nonumber \\
 &&\qquad 
   \times \left[ \delta^{(2)}\left({\bf \bar b}- {\bf r}_e\right) 
                - {\cal K}\bigl({\bf \bar b},0;{\bf r}_e,L|\mu\bigr)
                \right]
    \nonumber \\
 &&\qquad 
   \times \left[  \delta^{(2)}\left({\bf r}_e- {\bf b}\right)
                - {\cal K}\bigl({\bf r}_e,L;{\bf b},0|\mu\bigr) \right]\, .
  \label{4.25}
\end{eqnarray}   
The total photoabsorption cross section is the sum of both
\begin{eqnarray}
  &&\sigma_{\rm total}^{\gamma^*\to q\, \bar{q}}
  = N_c\, \alpha_{\rm em}\, \int d\alpha\, 
    \int {\it d}{\bf b}\, {\it d}{\bf \bar b}\, 
    \Phi\left({\bf b};{\bf \bar b};\alpha\right)
    \nonumber \\
 &&\qquad 
   \times \left[ 2\, \delta^{(2)}\left({\bf \bar b}- {\bf b}\right) 
                - {\cal K}\bigl({\bf \bar b},0;{\bf b},L|\mu\bigr)
                \right.
     \nonumber \\
 &&\qquad\qquad\qquad\qquad\qquad
    \left. - {\cal K}\bigl({\bf \bar b},L;{\bf b},0|\mu\bigr)
                \right]\, .
  \label{4.26}
\end{eqnarray}   
Before turning in the next subsection to the discussion of these
expressions, we remark shortly on a phase convention implicitly
used in the above results:

The first $(N-1)$ interactions in the above photoabsorption cross 
sections are contact terms. Contact terms stand either to the
right or to the left of the cut. The first interaction thus occurs 
for each non-vanishing contribution to (\ref{4.24}) at {\it different}
longitudinal positions $z_a$, $z_b$ in the amplitude and complex 
conjugate amplitude. As a consequence, the free incoming wavefunction
has to be evolved to these different positions in $M_{\rm fi}$ and 
$M_{\rm fi}^\dagger$. We explain in Appendix~\ref{appb} following 
(\ref{b.8}) that the expression $\Phi$ for the squared incoming 
wavefunction is only correct as long as $z_a = z_b$. If one
insists on using $\Phi$ for $z_a \not= z_b$, one is forced
to adopt the phase convention:
\begin{eqnarray}
  && \int {\it d}{\bf \bar b}\, 
    \Phi\left({\bf b};{\bf \bar b};\alpha\right)\, 
      {\cal K}_0\bigl({\bf \bar b},z_a;{\bf \bar r},z_b
       |\nu\alpha(1-\alpha)\bigr) 
    \nonumber \\
 && \qquad = \Phi\left({\bf b};{\bf \bar r};\alpha\right)\, 
    e^{-i\, q\, \left(z_b-z_a\right)}\, .
  \label{4.27}
\end{eqnarray}   
Strictly speaking, using $\Phi$ one has done the 
$y_L$- and $\bar y_L$-integrations in the photodissociation 
probability (\ref{3.7}) {\it before} specifying the true endpoints
of these integrations which are given by the positions
$z_a$, $z_b$ of the first interactions in $M_{\rm fi}$ and 
$M_{\rm fi}^\dagger$. The phase convention (\ref{4.27}) corrects
for the part of the $y_L$ integral which one misses in assuming
$z_a = z_b$. Here, we shortly illustrate
the consequences of (\ref{4.27}):

With the help of (\ref{4.27}), one checks immediately that the
$N=1$ contribution to (\ref{4.24}) agrees with the result (\ref{4.7})
of Nikolaev and Zakharov. There is no additional phase factor in
this case since $z_a = z_b$. However, there is an additional
phase factor for all contributions $N>1$. In the notationally 
simplest case, $N=2$, we find 
\begin{eqnarray}
  &&\sigma_{inel.}^{\gamma^*\to q\, \bar{q}}(N=2)
  \nonumber \\
  && \qquad = - N_c\, \alpha_{\rm em}\, {\rm Re}\, \int d\alpha\, 
    \int {\it d}\xi'\, n(\xi') \int_{\xi'} {\it d}\xi n(\xi)
  \nonumber \\
  && \qquad \qquad \times 
     \int {\it d}{\bf r}\, {\it d}{\bf \bar r}\, 
    \Phi\left({\bf r};{\bf \bar r};\alpha\right)\,
    e^{iq\left(\xi-\xi'\right)}  
    \nonumber \\
 && \qquad \qquad \times \, 
    \sigma\left({\bf \bar r}\right)\,
    {\cal K}_0\bigl({\bf \bar r},\xi';{\bf r},\xi
       |\mu\bigr)\, 
       \sigma\left({\bf r}\right)\, .
  \label{4.28}
\end{eqnarray}   
The phase factor $\exp\lbrace {iq\left(\xi-\xi'\right)}\rbrace$
is sensitive to the difference between the longitudinal position
of the first scattering center in $M_{\rm fi}$ and 
$M_{\rm fi}^\dagger$. The scale is set by the inverse coherence length
\begin{equation}
  q = \frac{Q^2}{2\nu} - \frac{m^2}{2\nu\alpha(1-\alpha)}
    = \frac{1}{l_f}\, .
    \label{4.29}
\end{equation}
%
%
It can be neglected in the high energy limit $\nu\to\infty$ where
the longitudinal extension of the target becomes small compared to
$1/q$.

\subsection{Colour triviality of the one-fold differential 
photoabsorption cross section} 
\label{sec4e}

Colour triviality of the inelastic, diffractive and total
photoabsorption cross section (\ref{4.24})-(\ref{4.26}) is the 
result of a complete
diagrammatic cancellation which arises due to the identities
in Fig.~\ref{fig7} (a) and (b). These identities are stronger than
implied by the optical theorem since they are based only on the
transverse momentum integration of one 
of the two quarks in the final state.
This makes it possible to use them in the simplification of
more differential observables. Here, we study the one-fold 
differential photoabsorption cross section
\begin{eqnarray}
 \frac{d\sigma^{\gamma^*\to q\bar{q}}}{d\alpha\,
   d{\bf p}_1^\perp}
 && =  \frac{\alpha_{\rm em}}{(2\pi)^2}
    \int {\it d}{\bf b}_1\,{\it d}{\bf b}_2\, 
    {\it d}{\bf \bar b}_1\, {\it d}{\bf \bar b}_2\,
    \Phi(\Delta {\bf b}; \Delta {\bf \bar b};\alpha)
      \nonumber \\
 && \quad \times 
    \int {\it d}{\bf x}_\perp\, {\it d}{\bf x'}_\perp\, 
      {\it d}{\bf \bar x'}_\perp\, 
      e^{i\, {\bf p}_1^\perp\cdot(
                {\bf x'}_\perp - {\bf \bar x'}_\perp)}
      \nonumber \\
  &&\quad \times 
    \Big\langle\, \bar G({\bf \bar b}_2;{\bf x}\vert p_2)
    \bar G({\bf x};{\bf b}_2\vert p_2)
      \nonumber \\
  &&\quad \times 
    \bar G({\bf b}_1;{\bf x'}\vert p_1)\,
    \bar G({\bf \bar x'};{\bf \bar b}_1\vert p_1)\Big\rangle\, .
  \label{4.30}
\end{eqnarray}
We shall derive for the inelastic part of (\ref{4.30}) an 
expression in terms of
\begin{equation}
 \bar\sigma({\bf b})  = 2\, C_F
      \int \frac{d\bbox{\kappa}_\perp}{(2\pi)^2}\, 
      \vert a_0(\bbox{\kappa}_\perp)\vert^2\,
      e^{-i\bbox{\kappa}_\perp\cdot{\bf b} }\, ,
  \label{4.31}
\end{equation}
which is closely related to the dipole cross section, 
\begin{equation}
  \sigma({\bf r}_1-{\bf \bar r}_1) = \bar\sigma(0)
       - \bar\sigma\left({\bf r}_1-{\bf \bar r}_1\right)\, .
  \label{4.32}
\end{equation}
The term $\bar\sigma(0)$ corresponds to a contact term for which 
both vertices are linked to the same quark of the 
$q$-$\bar{q}$-system. In intermediate steps of our calculation,
$\bar\sigma$ is a useful bookkeeping device to evaluate
the diagrams in Fig.~\ref{fig10}a. Our final result,
however, will depend only on the dipole cross section.

\subsubsection{Inelastic one-fold differential cross section}

\begin{figure}[h]\epsfxsize=8.7cm 
\centerline{\epsfbox{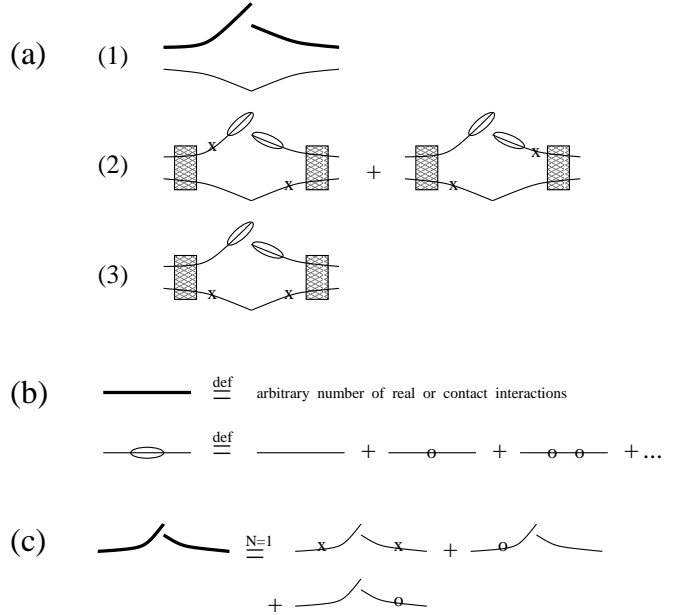}}
\vspace{0.5cm}
\caption{(a) The non-vanishing contributions to the inelastic
part of the one-fold differential photoabsorption cross 
section~(\protect\ref{4.30})
as classified in appendix~\protect\ref{appc}. (b) Definition
of diagrammatic shorthands used in (a). (c) Order $N=1$ contribution
to (a1).
}\label{fig10}
\end{figure}

The diagrammatic book-keeping of the inelastic contributions
to the one-fold differential cross section (\ref{4.30}) are
involved.
In appendix~\ref{appc}, we classify all contributing diagrams 
and we use the identities Fig.~\ref{fig7}a and b to
show that many of these diagrams cancel each other. As a result
of this analysis, the remaining non-vanishing contributions can 
be represented by the diagrams in Fig.~\ref{fig10}a. 
The colour trace for the $N$-th order terms of all 
these diagrams is $N_c\, C_F^N$: the 
inelastic part of the cross section (\ref{4.30}) is
thus free of colour interference terms, it is colour trivial.
%

We start with the $N=1$ contribution to the diagram Fig.~\ref{fig10}a1.
\begin{equation}
  e^{i{\bf p}_1^{\perp}\cdot\left({\bf x}'-\bar{\bf x}'\right)}\, 
  \int_0^L d\xi\, \frac{n(\xi)}{2}\, \left[ 
    \bar\sigma\left({\bf r}_1-{\bf \bar r}_1\right)
    - \bar\sigma(0) \right]\, 
     \Big\vert_{{\bf r}_1-{\bf \bar r}_1 = {\bf x}'-\bar{\bf x}'}\, .
  \label{4.33}
\end{equation}
This contribution is shown explicitly in Fig.~\ref{fig10}c. 
The relative transverse separation
${\bf r}_1-{\bf \bar r}_1$ between the quark in the amplitude 
and complex conjugated amplitude does not change with longitudinal
position $\xi$. This is a consequence of evolving on both sides
of the cut with Green's functions of same energy $p_1$. It makes
the iteration of the $N=1$ contribution to arbitrary high orders
particularly easy. The total contribution to Fig.~\ref{fig10}a1
reads
\begin{equation}
 e^{i{\bf p}_1^{\perp}\cdot\left({\bf x}'-\bar{\bf x}'\right)}\,
 \left(
   e^{-\int_0^L d\xi\, \frac{n(\xi)}{2}\, 
       \sigma\left({\bf x}'-\bar{\bf x}'\right)}
     - e^{-\int_0^L d\xi\, \frac{n(\xi)}{2}\, \bar\sigma\left(0\right)}
     \right)\, .
  \label{4.34}
\end{equation}
The first term is just the exponentiation of the $N=1$-
contribution. The second term subtracts those $N$-th order
contributions which involve only contact terms and thus do
not add to the inelastic contribution.

To evaluate the three contributions in Fig.~\ref{fig10}a2 and a3, 
we assume that the real momentum transfer occurs at an arbitrary 
but fixed longitudinal position $\check{\xi}$. The evolution of the
free incoming Fock state up to $\check{\xi}$ is then 
described by the path-integral (\ref{4.18}). The part of 
the expression for $\xi >\check{\xi}$ is given by 
\begin{eqnarray}
  && e^{i{\bf p}_1^{\perp}\cdot\left({\bf r}_1(\check{\xi})
      -\bar{\bf r}_1(\check{\xi})\right)}\,
  e^{-\int_{\check{\xi}}^L d\xi\, \frac{n(\xi)}{2}\, \bar\sigma\left(0\right)}
  \nonumber \\
  && \times \left[ \sigma\left(\bar{\bf r}_2(\check{\xi})
                    -\bar{\bf r}_1(\check{\xi})\right)
        + \sigma\left({\bf r}_2(\check{\xi})-{\bf r}_1(\check{\xi})\right)
        - \bar\sigma\left(0\right)
        \right]\, .
  \label{4.35}
\end{eqnarray}
Here, the real interactions at $\check{\xi}$ combine to the second
line of (\ref{4.35}) and the arbitrary number of contact terms 
at $\xi > \check{\xi}$ leads to the exponential of  $\bar{\sigma}(0)$.
Adding up all contributions of Fig.~\ref{fig10}a, we find for the
differential photoabsorption cross section (\ref{4.30}):
%
%
\begin{eqnarray}
 &&\frac{d\sigma^{\gamma^*\to q\bar{q}}_{\rm inel.}}{d\alpha\,
   d{\bf p}_1^\perp}
 \nonumber \\
 && =  \frac{\alpha_{\rm em}\, N_c}{(2\pi)^2}
    \int {\it d}{\bf b}\,{\it d}\bar{\bf b}\, 
    \Phi({\bf b};{\bf \bar b};\alpha)
      \nonumber \\
 && \quad \times 
 \left[ e^{i\, {\bf p}_1^\perp\cdot({\bf b} - {\bf \bar{b}})}\, 
   \left(e^{-\int_0^L d\xi\, \frac{n(\xi)}{2}\, 
       \sigma\left({\bf b}-\bar{\bf b}\right)}
     \right. \right.
   \nonumber \\
  && \qquad \qquad \qquad \qquad \qquad \left. \left. 
     - e^{-\int_0^L d\xi\, \frac{n(\xi)}{2}\, \bar\sigma\left(0\right)}
     \right) \right.    
      \nonumber \\
  &&\qquad \quad + \int_{\xi_{nl}}^L d\check{\xi}\, 
    \frac{n(\check{\xi})}{2} \int
    d{\bf r}\, d\bar{\bf r}\, 
    {\cal K}(\bar{\bf b},0;\bar{\bf r},\check{\xi}|\mu)\, 
      \nonumber \\
  &&\qquad \qquad \times 
  \lbrace \sigma(\bar{\bf r}) + \sigma({\bf r}) - \bar{\sigma}(0)\rbrace\, 
  {\cal K}({\bf r},\check{\xi};{\bf b},0|\mu)\, 
  \nonumber \\
  &&\qquad  \qquad \times \left.
    e^{-\int_{\check{\xi}}^L d\xi\, \frac{n(\xi)}{2}\, 
      \bar\sigma\left(0\right)}\,
    e^{i\, {\bf p}_1^\perp\cdot({\bf r} - {\bf \bar{r}})}
    \right] \, .
  \label{4.36}
\end{eqnarray}
The first term in the wide brackets stems from the contribution
Fig.~\ref{fig10}a1) given in (\ref{4.34}), the second term denotes the
diagrams Fig.~\ref{fig10}a2) and a3). Here, the Green's functions
${\cal K}$ describe the dynamical evolution of the $q$-$\bar{q}$ dipoles
up to the real interaction at $\check{\xi}$, from which point onwards
the dynamics is given by (\ref{4.35}). In analogy to (\ref{4.23}),
the lower boundary $\xi_{nl}$ of the intergration over $\check{\xi}$
is determined by the last interaction point before the real interaction.
To remove this indirectly defined variable from our solution, we use
again the opacity expansion (\ref{4.20}) to show that
\begin{eqnarray}
 &&\frac{d\sigma^{\gamma^*\to q\bar{q}}_{\rm inel.}}{d\alpha\,
   d{\bf p}_1^\perp}
  =  \frac{\alpha_{\rm em}\, N_c}{(2\pi)^2}
    \int {\it d}{\bf b}\,{\it d}\bar{\bf b}\, 
    \Phi({\bf b};{\bf \bar b};\alpha)
      \nonumber \\
 && \qquad \times 
 \left[ e^{i\, {\bf p}_1^\perp\cdot({\bf b} - {\bf \bar{b}})}\, 
   e^{-\int_0^L d\xi\, \frac{n(\xi)}{2}\, 
       \sigma\left({\bf b}-\bar{\bf b}\right)}
      - \int d{\bf r}\, d\bar{\bf r}
     \right.
   \nonumber \\
  &&\qquad \qquad \left. 
    e^{i\, {\bf p}_1^\perp\cdot({\bf r} - {\bf \bar{r}})}\, 
    {\cal K}(\bar{\bf b},0;\bar{\bf r},L|\mu)\, 
    {\cal K}({\bf r},L;{\bf b},0|\mu)\, 
    \right] \, .
  \label{4.37}
\end{eqnarray}
It is easy to check that the ${\bf p}_1^\perp$-integration of
this expression coincides with the inclusive inelastic photoabsorption
cross section (\ref{4.24}).

\subsubsection{Diffractive and total photoabsorption cross section
with one jet resolved}

The diffractive one-fold differential photoabsorption cross section
involves by definition only contact terms. Due to energy momentum
conservation, at least one contact term is required on the amplitude
level. An arbitrary non-vanishing number of contact terms in the
amplitude translates into a factor $\delta - {\cal K}$, see the
discussion of Fig.~\ref{fig7} (d). Hence, the diffractive contribution
to the cross section reads
\begin{eqnarray}
 \frac{d\sigma^{\gamma^*\to q\bar{q}}_{\rm diff.}}{d\alpha\,
   d{\bf p}_1^\perp}
 &=&  \frac{\alpha_{\rm em}\, N_c}{(2\pi)^2}
    \int {\it d}{\bf b}\,{\it d}\bar{\bf b}\, 
    \Phi({\bf b};{\bf \bar b};\alpha)
    \int d{\bf r}\, d\bar{\bf r}\, 
    e^{i\, {\bf p}_1^\perp\cdot({\bf r} - {\bf \bar{r}})}
      \nonumber \\
 && \qquad \times 
 \left[ \delta^{(2)}(\bar{\bf b}- \bar{\bf r})
        - {\cal K}(\bar{\bf b},0;\bar{\bf r},L|\mu) \right]
 \nonumber \\
 && \qquad \times 
 \left[ \delta^{(2)}({\bf b}- {\bf r})
        - {\cal K}({\bf r},L;{\bf b},0|\mu) \right]\, .
  \label{4.38}
\end{eqnarray}
The one-fold differential total cross section is given by the sum
of the contributions (\ref{4.37}) and (\ref{4.38}),
\begin{eqnarray}
 &&\frac{d\sigma^{\gamma^*\to q\bar{q}}_{\rm total}}{d\alpha\,
   d{\bf p}_1^\perp}
 =  \frac{\alpha_{\rm em}\, N_c}{(2\pi)^2}
    \int {\it d}{\bf b}\,{\it d}\bar{\bf b}\, 
    \Phi({\bf b};{\bf \bar b};\alpha)
      \nonumber \\
 && \qquad \times 
 \left[ e^{i\, {\bf p}_1^\perp\cdot({\bf b} - {\bf \bar{b}})}\, 
       \left( e^{-\int_0^L d\xi\, \frac{n(\xi)}{2}\, 
              \sigma\left({\bf b}-\bar{\bf b}\right)}
             + 1 \right) \right.
      \nonumber \\
 && \qquad  \qquad 
    - \int d{\bf r}\,   
      e^{i\, {\bf p}_1^\perp\cdot({\bf r} - {\bf \bar{b}})}\,
      {\cal K}({\bf r},L;{\bf b},0|\mu)
      \nonumber \\
 && \qquad  \qquad \left.
    - \int d{\bf \bar r}\,   
      e^{i\, {\bf p}_1^\perp\cdot({\bf b} - {\bf \bar{r}})}\,
      {\cal K}({\bf \bar b},0;{\bf \bar r},L|\mu) \right]\, .
  \label{4.39}
\end{eqnarray}
%

\subsubsection{Non-trivial colour interference in one-fold and two-fold
differential photoabsorption cross sections} 

Colour triviality of differential cross sections cannot be taken
for granted. It has to be established by the diagrammatic techniques
used in the above subsections. To illustrate this point, we give in
the following simple examples of untagged differential cross sections
which show colour interference effects.  We calculate the 
inelastic part of the $N=2$ photoabsorption cross section 
(\ref{3.9}) in the $\nu\to\infty$ limit in which the $q$-$\bar{q}$ state
propagates along straight lines through the nuclear target. For
a target of thickness $L$ and homogeneous density $n(\xi) = n_0$,
\begin{eqnarray}
 && \frac{d\sigma^{\gamma^*\to q\bar{q}}_{\rm inel.}(N=2)}{d\alpha\,
   d{\bf p}_1^\perp\, d{\bf p}_2^\perp}
 \nonumber \\
 && =  \frac{\alpha_{\rm em}}{(2\pi)^4}
    \int {\it d}{\bf b}_1\,{\it d}{\bf b}_2\, 
    {\it d}{\bf \bar b}_1\, {\it d}{\bf \bar b}_2\,
    \Phi({\bf b}_1-{\bf b}_2;{\bf \bar b}_1-{\bf \bar b}_2;\alpha)
      \nonumber \\
 && \quad \times 
      e^{i\, {\bf p}_1^\perp\cdot(
                {\bf b}_1 - {\bf \bar{b}}_1)}\, 
      e^{-i\, {\bf p}_2^\perp\cdot(
                {\bf b}_2 - {\bf \bar b}_2)}\, 
      \frac{n_0^2\, L^2}{2\, C_F^2} \frac{1}{4}
      \nonumber \\
  &&\quad \times 
    \left[ \bar{\sigma}({\bf b}_1 - {\bf \bar{b}}_1)
          + \bar{\sigma}({\bf b}_2 - {\bf \bar{b}}_2) \right.
      \nonumber \\
  && \qquad \left.  - \bar{\sigma}({\bf b}_1 - {\bf \bar{b}}_2)
          - \bar{\sigma}({\bf b}_2 - {\bf \bar{b}}_1) \right]
      \nonumber \\
  &&\quad \times 
    \left[ {\rm Tr}\lbrack aabb\rbrack\, B^{(1)}
           + {\rm Tr}\lbrack abab\rbrack\, B^{(2)} \right]\, ,
  \label{4.40}
\end{eqnarray}
where we have introduced the notational shorthands
\begin{eqnarray}
  B^{(1)} &=& \bar{\sigma}({\bf b}_1 - {\bf \bar{b}}_1)
          + \bar{\sigma}({\bf b}_2 - {\bf \bar{b}}_2)
          \nonumber \\
          &&
          + \bar{\sigma}({\bf \bar b}_1 - {\bf \bar{b}}_2)
          + \bar{\sigma}({\bf b}_2 - {\bf b}_1)
          - 4\, \bar{\sigma}(0)\, ,
  \label{4.41} \\
   B^{(2)} &=& \bar{\sigma}({\bf \bar{b}}_1 - {\bf \bar{b}}_2)
          + \bar{\sigma}({\bf b}_1 - {\bf b}_2)
          \nonumber \\
          &&- \bar{\sigma}({\bf b}_1 - {\bf \bar{b}}_2)
          - \bar{\sigma}({\bf b}_2 - {\bf b}_1) \, .
  \label{4.42}
\end{eqnarray}
For the cross section (\ref{4.40}), colour triviality is the 
condition that $B^{(2)}$ vanishes. However, $B^{(2)}\not= 0$,
and colour interference terms remain in (\ref{4.40}). In contrast,
the elastic part of the $N=2$ photoabsorption cross section is colour
trivial since it involves by construction only contact terms. Colour
triviality (non-triviality) of the inelastic part of the photoabsorption
cross section thus implies automatically the colour triviality 
(non-triviality) of the corresponding total cross section.

 In general, the more inclusive the cross section, the more likely
it is colour trivial. In this sense, tagged cross sections are
very exclusive since they require detailed knowledge of the
final state of the medium. On the other hand, calculating more
inclusive observables from (\ref{4.40}), one recovers colour
trivial examples. Especially, the ${\bf p}_{2\perp}$-integral 
(or ${\bf p}_{1\perp}$-integral) of (\ref{4.40}) turns out to
be colour trivial. This is expected, of course, since these integrated
versions of (\ref{4.40}) determine the $N=2$ term of (\ref{4.37}) in the
$\nu\to\infty$-limit. 

Based on these results, one may ask whether all one-fold differential 
photoabsorption cross sections are colour trivial. This is not the
case as can be seen e.g. from the inelastic parts of the 
one-fold differential cross sections
\begin{eqnarray}
  &&\frac{d\sigma^{\gamma^*\to q\bar{q}}_{\rm inel.}}{d\alpha\,
   d\left({\bf p}_1^\perp - {\bf p}_2^\perp\right)}\, ,
   \label{4.43} \\
  &&\frac{d\sigma^{\gamma^*\to q\bar{q}}_{\rm inel.}}{d\alpha\,
   d\left({\bf p}_1^\perp + {\bf p}_2^\perp\right)}\, .
   \label{4.44}
\end{eqnarray}
For these cross sections, the arguments of $\bar\sigma$ in
the expression (\ref{4.42}) for $B^{(2)}$ are constraint
by ${\bf b}_1 - {\bf \bar{b}}_1 = \pm \left( 
{\bf b}_2 - {\bf \bar b}_2\right)$. It is easy to check
that $B^{(2)}\not=0$, i.e., both results are not colour
trivial for $N=2$ and thus they cannot be colour trivial
for $N=$ arbitrary. For a colour trivial expression, it is
crucial that (\ref{3.9}) is not integrated over some average
or relative momentum but over the transverse
momentum of a final state particle. This is necessary for
using the identities of Fig.~\ref{fig7} (a) and (b).

\subsection{Mueller's dipole formulas as eikonal approximation}
\label{sec4f}

Explicit calculations of the $p_\perp$-integrated and one-fold differential
photoabsorption cross sections (\ref{4.24})-(\ref{4.26}) and
(\ref{4.37})-(\ref{4.40}) respectively require an explicit
representation of the path integral ${\cal K}$. This 
necessarily involves an approximation. In this section, we 
study the eikonal approximation which substitutes
the path integrals $\cal K$  by their ultrarelativistic limit
\begin{equation}
  \lim\limits_{\nu\to\infty}
  {\cal K}\bigl(L,{\bf r};0,{\bf b}|\mu\bigr) 
  = \delta^{(2)}\left({\bf b}-{\bf r}\right)\, S({\bf b})\, . 
  \label{4.45}
\end{equation}
We consider a homogeneous
density distribution $n(\xi)=n_0$ for a target of longitudinal
extension $L$ and we write for the Glauber-type suppression factor
\begin{equation}
  S({\bf b}) = e^{-\frac{1}{2}\, n\, L\, \sigma({\bf b})}\, .
  \label{4.46}
\end{equation}
Also, we neglect the phase factors 
$\exp\lbrace {iq\left(\xi-\xi'\right)}\rbrace$ in this 
$\nu\to\infty$-limit, see our discussion of (\ref{4.28}), (\ref{4.29})
above. In the ultrarelativistic limit, all photoabsorption cross sections
reduce to standard Glauber-type expressions:
\begin{eqnarray}  
  \lim\limits_{\nu\to\infty}\sigma_{inel.}^{\gamma^*\to q\, \bar{q}}
  &=& N_c\, \alpha_{\rm em}\, \int d\alpha\, 
    \int {\it d}{\bf b}\, 
    \Phi\left({\bf b};{\bf b};\alpha\right)
    \nonumber \\
  && \qquad \times \left[1 - S^2({\bf b})\right]\, ,
  \label{4.47}\\
  \lim\limits_{\nu\to\infty}\sigma_{diff.}^{\gamma^*\to q\, \bar{q}}
  &=& N_c\, \alpha_{\rm em}\, \int d\alpha\, 
    \int {\it d}{\bf b}\, 
    \Phi\left({\bf b};{\bf b};\alpha\right)
    \nonumber \\
  && \qquad \times \left[1 - S({\bf b})\right]^2\, ,
  \label{4.48}\\
  \lim\limits_{\nu\to\infty}\sigma_{total}^{\gamma^*\to q\, \bar{q}}
  &=& 2\, N_c\, \alpha_{\rm em}\, \int d\alpha\, 
    \int {\it d}{\bf b}\, 
    \Phi\left({\bf b};{\bf b};\alpha\right)
    \nonumber \\
  && \qquad \times \left[1 - S({\bf b})\right]\, .
  \label{4.49}
\end{eqnarray}   
This total photoabsorption cross section is Mueller's dipole 
formula first derived in Ref.~\cite{M90}. Also, the elastic
cross section is known to describe the diffractive contribution
to the deep inelastic scattering structure function $F_2$ in the
ultrarelativistic limit~\cite{KL99}. 

 The one-fold differential cross sections read in the eikonal limit
\begin{eqnarray}
 &&\lim\limits_{\nu\to\infty}
 \frac{d\sigma^{\gamma^*\to q\bar{q}}_{\rm inel.}}{d\alpha\,
   d{\bf p}_1^\perp}
 =  \frac{\alpha_{\rm em}\, N_c}{(2\pi)^2}
    \int {\it d}{\bf b}\,{\it d}\bar{\bf b}\, 
    \Phi({\bf b};{\bf \bar b};\alpha)
    e^{i\, {\bf p}_1^\perp\cdot({\bf b} - {\bf \bar{b}})}
      \nonumber \\
 && \qquad \qquad \times 
 \left[ S({\bf b} - {\bf \bar{b}}) 
        - S({\bf \bar{b}})\, S({\bf b})\right]\, ,
   \label{4.50}\\
 &&\lim\limits_{\nu\to\infty}
 \frac{d\sigma^{\gamma^*\to q\bar{q}}_{\rm diff.}}{d\alpha\,
   d{\bf p}_1^\perp}
 =  \frac{\alpha_{\rm em}\, N_c}{(2\pi)^2}
    \int {\it d}{\bf b}\,{\it d}\bar{\bf b}\, 
    \Phi({\bf b};{\bf \bar b};\alpha)
    e^{i\, {\bf p}_1^\perp\cdot({\bf b} - {\bf \bar{b}})}
      \nonumber \\
 && \qquad \qquad \times 
 \left[ 1 - S({\bf \bar{b}})\right]\,  
 \left[ 1 - S({\bf b})\right]\, ,
  \label{4.51} \\
 &&\lim\limits_{\nu\to\infty}
 \frac{d\sigma^{\gamma^*\to q\bar{q}}_{\rm total}}{d\alpha\,
   d{\bf p}_1^\perp}
 =  \frac{\alpha_{\rm em}\, N_c}{(2\pi)^2}
    \int {\it d}{\bf b}\,{\it d}\bar{\bf b}\, 
    \Phi({\bf b};{\bf \bar b};\alpha)
    e^{i\, {\bf p}_1^\perp\cdot({\bf b} - {\bf \bar{b}})}
      \nonumber \\
 && \qquad \qquad \times 
 \left[ 1 + S({\bf b} - {\bf \bar{b}}) 
       - S({\bf \bar{b}}) - S({\bf b})\right]\, .
  \label{4.52}
\end{eqnarray}
The elastic contribution (\ref{4.51}) has been obtained previously
in calculations of the diffractive component of DIS electron nucleon
scattering with one resolved jet in the final state~\cite{KL99}.
Furthermore, the total cross section (\ref{4.52}) was previously
obtained by Mueller (see eq. (27) of Ref.~\cite{M99}) from a 
one-loop calculation arguing then for its general validity. 
In passing, we recall Mueller's interpretation of the four
terms in (\ref{4.52}): for the case ${p_1^{\perp}}^2 \ll Q^2$,
the first two terms turn out to be of equal size while the last
two terms are negligible~\cite{M99}. Looking at the diagrammatic
contributions one concludes that in this limit the term proportional
$S({\bf b} - {\bf \bar{b}})$ gives the probability that a
quark which gets many random kicks is found with relatively small
transverse momentum. The term proportional to $1$, on the other hand,
corresponds to the case of no scattering and can be viewed as
the quantum mechanical shadow of the term proportional to
$S({\bf b} - {\bf \bar{b}})$~\cite{M99}.

\subsection{Corrections to the eikonal approximation}
\label{sec4g}

To go beyond the eikonal limit, we discuss now the Gaussian dipole 
approximation for the path-integrals ${\cal K}$. It is based on the 
observation that the main 
support in the path integral (\ref{4.19}) comes from small
transverse distances $r = |{\bf r}|$, where the cross section
$\sigma({\bf r})$ in (\ref{4.8}) has a leading
quadratic dependence:
\begin{eqnarray}
  \sigma({\bf r}) \approx C(r)\, r^2\, .
  \label{4.53}
\end{eqnarray}
In explicit model calculations, one 
finds~\cite{Z96,WG99} that the $r$-dependence of $C(r)$ is a 
slow logarithmic one, and can be neglected. For the dipole cross
section in (\ref{4.8}), an expansion to order $r^2$ confirms
this feature. One finds 
\begin{equation}
  C = \frac{C_F}{2} \int^{\kappa_c} \frac{d^2\bbox{\kappa}_\perp}{(2\pi)^2}\, 
      \bbox{\kappa}_\perp^2\, |a_0(\bbox{\kappa}_\perp)|^2\, ,
  \label{4.54}
\end{equation}
where the $\bbox{\kappa}_\perp$-integral depends logarithmically
on the ultraviolet cut-off $\kappa_c$ which one has to introduce
in this approximation. We note that
$C$ provides a measure of the average transverse momentum transfer 
$\langle \bbox{\kappa}_\perp^2\rangle$ in a single scattering.
For sufficiently small $r$, when the 
$r$-dependence of $C$ can be neglected, the path integral 
(\ref{4.19}) reduces to that of a harmonic oscillator~\cite{Z96}
\begin{eqnarray}
  {\cal K}_{\rm osz}\bigl({\bf r}_2,L;{\bf r}_1,0|\mu\bigr) 
     &=& {A\over \pi\, i} \exp\left\{iAB({\bf r}_1^2 + {\bf r}_2^2)
       \right.
  \nonumber \\
  &&  \qquad \left.
             -2\,i\,A\,{\bf r}_1\cdot{\bf r}_2 \right\}\, ,
  \label{4.55} \\
  A &=&  {\mu\Omega\over 2\, \sin(\Omega\, L)}\, ,
  \label{4.56} \\
  B &=& \cos(\Omega\, L)\, ,
  \label{4.57}
\end{eqnarray}  
with the oscillator frequency
\begin{equation}
  \Omega = \frac{1-i}{\sqrt{2}}\, \sqrt{n_0\, C\over \mu}\, .
  \label{4.58}
\end{equation}
For what follows, it is important that the $\mu\to\infty$-limit 
of ${\cal K}_{\rm osz}$
coincides with the eikonal limit (\ref{4.45}) of the unapproximated
path-integral. To see this, we expand the norm of 
${\cal K}_{\rm osz}$ to leading order in $\mu$ and the phase to
next to leading order, 
\begin{eqnarray}
  &&{\cal K}_{\rm osz}\bigl({\bf r}_2,L;{\bf r}_1,0|\mu\bigr) =
  {{\mu + O\left(\mu^0\right)}\over 2\pi\, i\, L}\,
  \exp\left\{ {i\mu\over 2L}\left({\bf r}_e-{\bf r}\right)^2 \right.
  \nonumber \\
  && \qquad \qquad 
    \left.
    - {Ln_0C\over 6}\left[{\bf r}_e^2 + {\bf r}^2 
                           + {\bf r}_e\cdot{\bf r}\right]
    + O\left(\mu^{-1}\right) \right\}
  \nonumber \\
  && {\buildrel {\mu\to\infty}\over \longrightarrow} \, \qquad 
  \delta^{(2)}\left({\bf r}_e - {\bf r}\right)\, 
  e^{-\frac{1}{2}\, n_0\, L\, C\, \bbox{r}^2 }\, ,
  \label{4.59}
\end{eqnarray}
which coincides with (\ref{4.45}) for the quadratic ansatz
$\sigma\left({\bf r}\right) = C\, {\bf r}^2$. 
The harmonic oscillator approximation thus provides
an explicit representation for the path-integral ${\cal K}$ which
preserves the correct high energy limit. At least numerically, 
this allows for an explicit study of the $1/\mu$-corrections to the
eikonal limit of the cross sections (\ref{4.24})-(\ref{4.26}) and
(\ref{4.37})-(\ref{4.39}). To explore some qualitative features
of these $1/\mu$-corrections analytically, we take here recourse
to a Gaussian approximation $\Phi_G$ of the incoming Born wavefunction:
\begin{eqnarray}
  \Phi_G({\bf r},{\bf \bar r};\alpha) &=& 
            \psi({\bf r})\, \psi^*({\bf \bar r})\, ,
  \label{4.60} \\
  \psi({\bf r}) &=& \textstyle{1\over \pi\, R^2}
                \exp\left( -{\bf r}^2/R^2\right)\, ,
  \label{4.61} \\
  \frac{1}{R^2} &=& (1-\alpha)\alpha Q^2 + m^2\, ,
  \label{4.62}
\end{eqnarray}
where the radius $R$ is chosen to reproduce the characteristic
width of the fall-off of the exact Born probability $\Phi$
given in (\ref{b.8}) and (\ref{b.14}).

The time-evolved final state wavefunction reads
\begin{eqnarray}
  \Psi_f({\bf \bar r}) &=& \int d{\bf r}\,
   {\cal K}_{\rm osz}({\bf \bar r},L;{\bf r},0|\mu)\, \psi({\bf r})
   \nonumber\\
   &=& \frac{1 + O(\mu^{-1})}{\pi\, R^2}
       \exp\left[ -\frac{{\bf \bar r}^2}{R^2} - \frac{1}{2}
                  n_0\, L\, C\, {\bf \bar r}^2 + i\, q\, L
                  \right.
   \nonumber \\
     && \left. \qquad \qquad i\frac{{\bf \bar r}^2}{\mu}\, c_1
            - \frac{{\bf \bar r}^2}{\mu^2}\, c_2 + O(\mu^{-3})\right]\, .
 \label{4.63}
\end{eqnarray}
Here, we have used the phase convention (\ref{4.27}) to obtain the
term $i\, q\, L$ in the exponent. The shorthands $c_1$ and $c_2$
denote the leading real and imaginary $1/\mu$-corrections to the
final state wavefunction:
\begin{eqnarray}
  c_1 &=& \frac{L}{6} \left[ n_0^2L^2C^2 + 6\frac{n_0LC}{R^2}
                             + \frac{12}{R^4} \right]\, ,
      \label{4.64}\\
  c_2 &=& \frac{L^2}{15} \left[ n_0^3L^3C^3 + 10 
            \frac{n_0^2L^2C^2}{R^2} + 40 \frac{n_0LC}{R^4}
            + \frac{60}{R^6}\right]\, .
      \label{4.65}
\end{eqnarray}
Deviations from the eikonal limit can be calculated with the help
of (\ref{4.63}). For example, we find for the total photoabsorption
cross section (\ref{4.26}):
\begin{eqnarray}
  &&\sigma_{\rm total}^{\gamma^*\to q\bar{q}} - 
  \lim_{\nu \to \infty} \sigma_{\rm total}^{\gamma^*\to q\bar{q}}
  \nonumber \\
  &&\qquad = N_c\, \alpha_{\rm em}\, \int d\alpha\int d{\bf r}\, 
      \Phi_G({\bf r},{\bf r};\alpha)\, S({\bf r})
      \nonumber \\
  && \qquad \qquad \times 2\, \left[ 1 - e^{-{\bf r}^2c_2/\mu^2}
                     \cos\left(\frac{{\bf r}^2}{\mu}c_1 
                        + q\, L \right) \right]\, .
  \label{4.66}
\end{eqnarray}
In the $\nu\to\infty$-limit, this difference vanishes by construction.
To understand which scales determine the deviations from the eikonal
limit, we recall that $1/(\mu\, R^2) = q$ has an interpretation as
inverse coherence length $1/l_f$ and $n_0LC$ is the total transverse
momentum squared accumulated during the rescattering over a distance
$L$. The eikonal approximation is thus justified if the phases
determined by the coherence length and the total transverse 
energy $E_\perp^{\rm tot}$ are negligible:
\begin{eqnarray}
  q\, L &\ll& 1\, ,
  \label{4.67}\\
  E_\perp^{\rm tot}\, L = \frac{n_0LC}{2\mu}\, L &\ll& 1\, .
  \label{4.68}
\end{eqnarray}
We conclude this section with two technical remarks:

1. The explicit form of the incoming $q$-$\bar{q}$ wavefunctions
(\ref{b.6}), (\ref{b.13}) is given in terms of the Bessel
function $K_0$ which has an integral representation in terms 
of a Gaussian in ${\bf r}$, 
\begin{equation}
  K_0(\epsilon\, |{\bf r}|) = \frac{1}{2} \int_{-\infty}^0
  \frac{dy_L}{y_L}\, e^{-i \frac{\epsilon^2}{4\, y_L}{\bf r}^2
              + i\, y_L}\, .
  \label{4.69}
\end{equation}
The evolution of the Gaussian wavefunction $\psi({\bf r})$ in 
(\ref{4.63}) can thus be used for the exact calculation if 
$1/R^2$ is taken to be the width $\frac{\epsilon^2}{4\, y_L}$ 
of the integrand of (\ref{4.69}) and the $y_L$-integration is
done afterwards. However, to order $1/\mu$, terms proportional
to $1/R^4$ appear in (\ref{4.63}) and thus the $y_L$-integration
cannot be done analytically if $1/\mu$-corrections are included.
Only the $\mu\to\infty$-limit can be accessed analytically in this
way.

2. A surprising difficulty occurs in the calculation of the 
photoabsorption cross section (\ref{4.24}) from the harmonic
oscillator approximation (\ref{4.55}) if one tries to do the
${\bf r}_e$-integration first:
\begin{equation}
  F_{\rm osz}({\bf \bar r},{\bf r}) =
  \int d{\bf r}_e\, 
      {\cal K}_{\rm osz}\bigl({\bf \bar r},0;{\bf r}_e,L|\mu\bigr)\, 
    {\cal K}_{\rm osz}\bigl({\bf r}_e,L;{\bf r},0|\mu\bigr)\, .
  \label{4.70}
\end{equation} 
After a lengthy but straightforward calculation, we find for the 
$\mu\to\infty$-limit 
\begin{equation}
 \lim_{\mu\to\infty} F_{\rm osz}({\bf \bar r},{\bf r})
 = \delta^{(2)}\left( {\bf \bar r} - {\bf r}\right)\, 
  e^{-\frac{1}{4}\, n_0\, L\, C\, {\bf r}^2 }\, ,
    \label{4.71}
\end{equation}
whose exponent differs from that of the eikonal dipole formula 
(\ref{4.45}) by a factor $\frac{1}{4}$. This mismatch is an 
artefact stemming from a calculation which performs simplifications
on the cross section level before completing the dynamical evolution
on the amplitude level. Starting from (\ref{4.63}) for $\Psi_f({\bf r}_e)$
and calculating $\Psi_f({\bf r}_e)\, \Psi_f^*({\bf r}_e)$, one obviously
reproduces for (\ref{4.24}) the eikonal expression (\ref{4.47}) in the
$\mu\to\infty$-limit.

\section{Photodissociation via non-abelian Stokes's theorem} 
\label{sec5}

In this section, we argue that the dipole cross section (\ref{4.8})
parametrizes the transverse components of the chromoelectric field 
strength correlations $\langle F\, F\rangle$ in the nuclear medium. 
This suggests an interpretation of 
$\sigma({\bf r})$ which holds model-independent, i.e., irrespective 
of whether we have build up (\ref{4.8}) from the model ansatz 
(\ref{2.10}), or from some other model parametrization of the colour 
target field.

We first recall the non-abelian Stokes's theorem. This relates the 
integral of the non-abelian vector potential $A_\mu$ along a closed 
contour $C$ to the area integral of the field strength
tensor $F^{\mu\nu}$,~\cite{C75,A80,FGK81}
\begin{eqnarray}
  && {\rm Tr}\, {\cal P}\, \exp\left( \oint_C dz_\mu\,
    A^{\mu}(z)\right)
  \nonumber \\
  && = {\rm Tr}\, {\cal P}_{\rm area}\,
    \exp\left( \int d\sigma_{\mu\nu}(y)\, 
      U_{Ay}\, F^{\mu\nu}(y)\, U_{yA}\right)\, .
  \label{5.1}
\end{eqnarray}
Here, the $U_{Ay}$ are parallel transporters.
They ensure that the r.h.s. is gauge-invariant. Equation
(\ref{5.1}) does not depend on the choice of the position $A$,
since cyclicity of the trace allows to change the reference
point by substituting $U_{Ay} \to U_{BA}\, U_{Ay}$. 
The area ordering ${\cal P}_{\rm area}$
is defined by disecting the area enclosed by $C$ in many (ultimately: 
infinitessimally small) areas in such an order that the parallel 
transporters $C_i$ along the infinitessimal areas combine to $C$. 
As a consequence of the non-Abelian Gauss theorem, (\ref{5.1})
turns out to be independent of the orientation of the surface 
$d\sigma_{\mu\nu}(y)$~\cite{FGK81}. 

The total photoabsorption cross section (\ref{4.1}) is closely
related to the Stokes's theorem, as can be seen from the
representation
\begin{eqnarray}
  \sigma_{total}^{\gamma^*\to q\, \bar{q}}
  &=& \alpha_{\rm em}\, \int d\alpha\, 
    \int {\it d}{\bf b}_1\, {\it d}{\bf b}_2\, 
    {\it d}{\bf \bar b}_1\, {\it d}{\bf \bar b}_2\,
    \Phi(\Delta {\bf b}; \Delta {\bf \bar b};\alpha)
    \nonumber \\
 && \times 
    \int {\cal D}r_1\,  {\cal D}r_2\,  
    e^{\frac{i\nu}{2}\int \left(\alpha \dot{r}_2^2 +
        (1-\alpha) \dot{r}_1^2 \right)}
    \nonumber \\
 && \times 
    \int {\cal D}\bar{r}_1\,  {\cal D}\bar{r}_2\,
    e^{\frac{i\nu}{2}\int \left( - \alpha \dot{\bar{r}}_2^2
        - (1-\alpha) \dot{\bar{r}}_1^2 \right)}
    \nonumber \\
 && \times 
     \Big\langle {\rm Tr}\, {\cal P}\, 
     \exp\left( \oint_{C(r_1,r_2,\bar{r}_1,\bar{r}_2)} dz_\mu\,
    A^{\mu}(z)\right) \Big\rangle
  \label{5.2}
\end{eqnarray}   
Here, $C$ is defined by the transverse paths
${\bf r}_1(\xi)$, ${\bf \bar{r}}_1(\xi)$,
${\bf r}_2(\xi)$, ${\bf \bar{r}}_2(\xi)$ which denote the positions
of the rescattering quark and anti-quark in the path integral 
representation (\ref{2.17}) of the corresponding Green's functions. We 
can consider $C$ as a closed path since we work for a static potential 
$A_\mu = \delta_{\mu 0}\, A_0$ for which the Wilson line along 
purely transverse directions $d{\bf x}_\perp\, A_\perp(x)$ vanishes. 
The photoabsorption cross section (\ref{5.2}) is thus given by a 
closed Wilson loop which sums up the gluon field strength
of the medium in the area determined by the paths of the
rescattering $q$- and $\bar{q}$- quarks.

\begin{figure}[h]\epsfxsize=5.7cm 
\centerline{\epsfbox{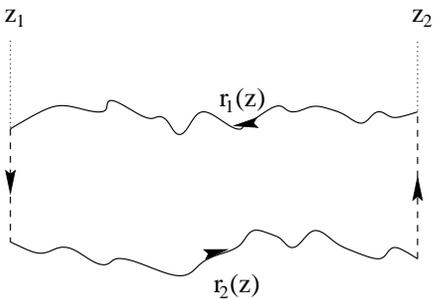}}
\vspace{0.5cm}
\caption{
The closed Wilson loop between longitudinal 
positions $z_1$ and $z_2$, as considered in (\protect\ref{5.3}). 
}\label{fig11}
\end{figure}

The connection between total hadronic cross sections as (\ref{5.2})
and closed Wilson loops has been studied extensively, see e.g.
\cite{N91,DFK94}. Phenomenological applications of the non-abelian
Stokes's theorem (\ref{5.1}) were pioneered by Dosch and 
coauthors~\cite{DS88,DFK94}. In Dosch's approach, one takes recourse
to an axial gauge in which the parallel transporters on the r.h.s.
of (\ref{5.1}) reduce to unit operators. The remaining expression
is then expanded in powers of $F$. 
Our discussion of (\ref{5.2}) borrows from this strategy. Clearly,
we cannot choose an axial gauge to remove the parallel transporters
in (\ref{5.1}), since our gauge freedom is already exhausted by
the definition (\ref{2.10}). However, we can invoke the proof of
colour triviality of (\ref{5.2}) to discuss the equivalent 
abelian problem with a rescaled vector potential $A_\mu \to
\sqrt{C_F}\, A_\mu$. We find
\begin{eqnarray}
     &&\Big\langle  \exp\left( 
       -i\, g\, \sqrt{C_F} 
       \oint_C\, ds\, A^{\mu}(\omega_s)\, 
       \frac{\partial \omega_s^{\mu}}{\partial s}\right) \Big\rangle
     \nonumber \\
     && = \Big\langle  \exp\left( 
       i\, g\, \sqrt{C_F} 
       \int_S\, F_{\mu\nu}(\omega_s)\, 
       \frac{\partial \omega_s^{\mu}}{\partial s}\,
       \frac{\partial \omega_s^{\nu}}{\partial x_\alpha}\,
       ds\, dx_\alpha
       \right) \Big\rangle
     \nonumber \\
     && = \Big\langle  \exp\left( 
       - i\, g\, \sqrt{C_F} 
       \int_{z_1}^{z_2}\, d\xi\,
       \int_{{\bf r}_1(\xi)}^{{\bf r}_2(\xi)} 
       dx^i_\perp\, E_i^\perp({\bf x}_\perp,\xi)
       \right) \Big\rangle
     \nonumber \\
     && = 1 - \int_{z_1}^{z_2}\, d\xi\, n(\xi)\, \frac{g^2\, C_F}{2}
     \Big\langle \int d\bar{\xi} \int_{{\bf r}_1(\xi)}^{{\bf r}_2(\xi)} 
     dx^i_\perp\, dx'^j_\perp
     \nonumber \\
     && \qquad \qquad \times
     E_i^\perp({\bf x}_\perp,\xi)\, 
     E_j^\perp({\bf x'}_\perp,\xi+\bar{\xi})\,
     \Big\rangle + O\left(n^2(\xi)\right)\, .
  \label{5.3}
\end{eqnarray}  
Here, $\omega_s^{\mu}$ denotes the path around the contour $C$, and 
the notation is specified in Fig.~\ref{fig11}. 
%
We have used the 
particular form of a static colour potential 
$A_\mu = \delta_{\mu 0}\, A_0$ to rewrite the r.h.s. of (\ref{5.1})
in terms of transverse electric fields. The detailed arguments for the 
shift of integration variables needed to arrive at the last equation
of (\ref{5.3}) are given in appendix B of Ref.~\cite{WG99}. The same
appendix contains the derivation of the dipole cross section (\ref{4.8})
from a closed abelian Wilson loop:
\begin{eqnarray}
     &&\Big\langle  \exp\left( 
       -i\, g\, \sqrt{C_F} 
       \oint_C\, ds\, A^{\mu}(\omega_s)\, 
       \frac{\partial \omega_s^{\mu}}{\partial s}\right) \Big\rangle\, =
     \nonumber \\
     &&\Big\langle  \exp\left( 
       -i\, g\, \sqrt{C_F} 
       \int_{z_1}^{z_2}\, d\xi\, 
       \left[ A_0({\bf r}_1(\xi),\xi) -  A_0({\bf r}_2(\xi),\xi)\right]
       \right) \Big\rangle
     \nonumber \\
     && = \exp\left( - \int_{z_1}^{z_2}\, d\xi\, n(\xi)\, 
       \sigma\left({\bf r}_1(\xi) - {\bf r}_2(\xi)\right) 
       \right)\, .
  \label{5.4}
\end{eqnarray}  
Comparing (\ref{5.4}) to (\ref{5.3}), we can relate the dipole
cross section to the two-point correlation function of the transverse
electric field strength,
\begin{eqnarray}
   &&\sigma\left({\bf r}_1(\xi) - {\bf r}_2(\xi)\right)
   = \Big\langle \int d\bar{\xi} \int_{{\bf r}_1(\xi)}^{{\bf r}_2(\xi)} 
     dx^i_\perp\, dx'^j_\perp
     \nonumber \\
     && \qquad \qquad \qquad \times
     E_i^\perp({\bf x}_\perp,\xi)\, 
     E_j^\perp({\bf x'}_\perp,\xi+\bar{\xi})\,
     \Big\rangle\, .
     \label{5.5}
\end{eqnarray}
As long as we parametrize this dipole cross section by the Gaussian 
approximation $\sigma({\bf r}) = C\, r^2$, only the average
of the colour field strength enters our final results in form of 
one single parameter $C$. The final result is model-independent
in the sense that it does not depend on the model-specific way in 
which the colour field strength giving rise to $C$ was modelled.

\section{Conclusion}
\label{sec6}

Colour triviality provides a crucial simplification of an otherwise
untractable problem. In general, the $N$-fold rescattering of a 
parton in a colour target field leads in the cross section 
to colour traces over $2N$ generators. For sufficiently exclusive 
processes, the huge number of colour interference terms 
thus limits explicit calculations of the soft nuclear 
dependence of hard partonic processes to the case of very few 
rescatterings~\cite{GLV99}. Colour triviality is the consequence of 
a complete diagrammatic cancellation between these different colour 
interference terms. For colour trivial observables, all non-abelian 
complications reduce to the rescaling of coupling constants by 
appropriate powers of Casimirs. The multiple scattering problem 
thus becomes an abelian one. The soft nuclear dependence of hard 
colour-trivial observables can be described in terms of a QCD 
dipole cross section which absorbs the leading medium-dependence 
in a one-parameter estimate $C$ of the average target colour field 
strength. This makes the identification of colour-trivial observables 
of particular interest for relativistic heavy ion collisions at 
RHIC and LHC where very little is known about the medium a priori 
but many ``hard probes'' are expected to receive sizeable nuclear 
modifications due to rescattering effects.

In the present work, we have derived an explicit general expression 
for the rescattering effects on a hard coloured parton inside a 
spatially extended nuclear medium. We have then studied in detail
for the simple example of virtual photodissociation under which 
conditions this general rescattering formula results in colour-trivial
observables which can be parametrized by a QCD dipole cross section.
In detail:

The non-abelian Furry wavefunction (\ref{2.12}) derived in 
section~\ref{sec2} is a high-energy approximation to the solution 
of the non-abelian Dirac equation in a spatially extended colour
field. In contrast to other approximate solutions~\cite{BH96},
it is accurate up to order $O(1/E^2)$ 
in the phase. This is known to be indespensable for
calculating the nuclear dependence of observables which are 
determined by the destructive interference between different 
production amplitudes as e.g. the LPM-effect or the nuclear 
dependence of Drell-Yan yields. The Furry
wavefunction (\ref{2.12}) thus provides a unified starting point
for the description of the nuclear dependence of a large class
of observables. 

Based on the Green's function $\bar{G}$ which describes the 
dynamical evolution of the non-abelian Furry wavefunction, we
have derived in section~\ref{sec4} a set of diagrammatic identities
which play the key role in proofs of colour triviality. These
identities exploit the integration over the transverse
momentum of a single final
state particle only. They are thus much stronger than diagrammatic
cancellations implied by the optical theorem. As a consequence,
they ensure for the example of the $\gamma^*\to q\bar{q}$ 
process a colour trivial result not only for the 
total, elastic and inelastic inclusive photoabsorption cross section
but also for cases in which one jet is resolved in the final state.
All these observables can be described in terms of the same
one-parameter QCD dipole cross section.

Earlier applications of these diagrammatic techniques as well
as statements about the colour triviality of 
the photoabsorption cross section exist 
~\cite{M90,BDMPS2,BDMPS3,BDMS,BDMS-Zak,KM98,M99,KL99}. 
Here, we have given relatively short and complete
proofs of these statements by exploiting the advantages
of a new and compact configuration-space notation implied 
by the Furry wavefunction. In contrast to previous discussions, our 
formulation includes the transverse dynamical evolution of the partons. 
For the example of the photoabsorption cross section studied here,
this allowed us to quantify the leading deviations from the
eikonal limit which grow at fixed energy proportional to $L^2$. 
Moreover, the compactness of our notation made
it possible to discuss in detail technically rather involved
processes as e.g. the inelastic and total photoabsorption
cross section with one jet resolved in the final state. This points to the 
strength of the present approach which we believe to be suited for the
explicit discussion of more complicated processes as, e.g., further 
studies of the transverse momentum dependence of the non-abelian 
LPM-effect~\cite{BDMS,WG99}
or photodissociation including initial state gluon radiation~\cite{BGH99}. 

\acknowledgements
I am indebted to Miklos Gyulassy for many helpful and inspiring 
discussions, a critical reading of this manuscript, and  
the hospitality extended to me at Columbia University. I thank Al 
Mueller for several discussions about "contact terms" which have 
strongly influenced sections~\ref{sec4d} - ~\ref{sec4f}.
Thanks go to Yuri Kovchegov for his patience in explaining to me many
technical details of Refs.~\cite{KL99,KM98}. Helpful comments by 
E. Iancu, P. Levai, E. Levin, A. Leonidov, L. McLerran, R. Venugopalan 
and I. Vitev are gratefully acknowledged. This work was supported by a 
CERN Fellowship and by the Director, Office of Energy Research, Division of
Nuclear Physics of the Office of High Energy and Nuclear Physics of the
U.S. Department of Energy under Contract No. De-FG-02-92ER-40764.

\appendix
\section{Non-abelian Furry approximation}
\label{appa}

In this appendix, we derive the non-abelian Furry wavefunction
(\ref{2.12}) from the set of $N$-scattering Feynman diagrams (\ref{2.11}).
For an $N$-fold rescattering diagram, we use the notation
${\bf p}_{N+1} \equiv {\bf p}$ for the final state momentum.
To simplify $I^{(N)}({\bf y})$ in (\ref{2.11}),  
we do the longitudinal momentum integrals by 
contour integration
\begin{eqnarray}
  &&\int {dp_{i}^L\over (2\pi)}\,
   {i\, (\not{p}_{i} + m)\, \gamma_0\over
                     {p_{i}^2 - m^2 + i\,\epsilon}}\,
   e^{i\,p_{i}^L\, (x_i^L - x_{i-1}^L)}\nonumber \\
   && \qquad
   = {(\not{p}_{i} + m)\, \gamma_0\over 2\, p_{i}^L}
    \Theta(x_i^L - x_{i-1}^L)\, e^{i\,p_{i}^L\, (x_i^L - x_{i-1}^L)}\, .
  \label{a.1}
\end{eqnarray}
We note that, strictly speaking, the position variables $x_i$ in this 
expression are integration variables and do not coincide with the
center of the $i$-th scattering potential. A more detailed analysis
of (\ref{2.11}) involves for a particular model also the study of the 
$p^L_i$-poles of the Fourier-transformed single scattering potentials. 
For a Yukawa-type potential $1/[({\bf p}_i - {\bf p}_{i-1})^2 + M^2]$, 
e.g., these poles give additional contributions to the $p^L_i$-integrals, 
which are however exponentially suppressed due to the Debye-screening 
mass $M$. These details are discussed explicitly in  
Refs.~\cite{GW94,BDMPS2} and for contact terms in Ref.~\cite{GLV2}. 
In the end, the only $O(1/E)$-contribution
to the $p_i^L$-integration turns out to be given by (\ref{a.1}). 
On the r.h.s. of this equation, $p_{i}^L$ is determined by the pole
value to order $O(1/E)$, $p_i^L = p\, - 
\textstyle{{p_i^{\perp}}^2\over 2\, p}$ where $p^2 = E^2 - m^2$.
Using ${\bf x}_0 \equiv {\bf y}$, we can rewrite (\ref{2.11}) as
%
%
\begin{eqnarray}
  &&I^{(N)}({\bf y}) =
  e^{-\, i\, p\, y_L}\,
  {\cal P}\, \left(
  \prod_{i=1}^N\int {d^2{\bf p}_i^\perp\over (2\pi)^2}\, d^3{\bf x}_i\,
                    \Theta(x_i^L - x_{i-1}^L)\right.
                  \nonumber \\
   &&\qquad \left.\times {(\not{p}_i + m)\, \gamma_0\over
                     {2\, p_i^L}}\, \lbrack -i\, A_0({\bf x}_i)
                    \rbrack\, 
                    e^{i\, {\bf p}_i^{\perp}\cdot 
                      ({\bf x}_i^\perp - {\bf x}_{i-1}^\perp)}
                    \right.
                    \nonumber \\
   &&\qquad \left. \times  e^{- i{{p_i^{\perp}}^2\over 2\, p}
                       (x_i^L - x_{i-1}^L)}\, \right)\, 
                    e^{-i{\bf p}\cdot{\bf x}_N
                       + i{p\, x_N^L}}\,  
                     v^{(r)}({\bf p})\, .
  \label{a.2}
\end{eqnarray}
We now consider the spinor structure of this
expression. To order $O(1/E^2)$, each quark propagator introduces
a numerator
\begin{equation}
  (\not{p}_i + m)\gamma_0 \approx E(\gamma_0-\gamma_3)\gamma_0
              - {\bf \gamma}^\perp\cdot{\bf p}^{\perp}_i\, \gamma_0
             + \gamma_3\, \gamma_0\,
                 \textstyle{{p_i^{\perp}}^2\over2\, p}\, .
  \label{a.3}
\end{equation}
The normalization of $I^{(N)}({\bf y})$ is needed to leading
order in energy only. This allows us to neglect the mass term
in (\ref{a.3}). For the same reason, it is sufficient to keep only the leading
term $E(\gamma_0-\gamma_3)\gamma_0$ for all numerators (\ref{a.3})
with $i \geq 2$. Further simplification of (\ref{a.2}) is then
possible, using
\begin{equation}
    \left(E(\gamma_0-\gamma_3)\gamma_0\right)^n 
  = 2^{n-1}\, E^n\, (\gamma_0-\gamma_3)\gamma_0
  \label{a.4}\, .
\end{equation}
The case $i=1$ is different: depending on the
explicit form of the production vertex, the leading order contribution
of this numerator can cancel. We have to keep the numerator
$(\not{p}_1 + m)\gamma_0$ to order $O(1/E)$.
To this aim, we substitute in (\ref{a.2})
\begin{equation}
  (\not{p}_1 + m)\gamma_0 \longrightarrow
  E(\gamma_0-\gamma_3)\gamma_0 - i{\bbox{\gamma}}\cdot
  {\partial\over \partial {\bf y}}\gamma_0\, ,
  \label{a.5}
\end{equation}
where the differential operator acts on the big bracket in 
(\ref{a.2}). Equation (\ref{a.5}) is equivalent to (\ref{a.3})
with the transverse components written in configuration space.
In a coordinate system with ${\bf p} \parallel {\bf n}$, the
above equation takes the form
\begin{eqnarray}
  \not{p}_1\gamma_0 &=& p(\gamma_0-\gamma_3)\gamma_0
  + \bbox{\alpha}^\perp\cdot({\bf p}_1^\perp - {\bf p}^\perp)
  \nonumber \\
  && - \alpha^L\, \left( {{{\bf p}_1^\perp}^2\over 2\, p} -
                   {{{\bf p}^\perp}^2\over 2\, p} \right) 
  \nonumber \\
  \longrightarrow && 
  p\, (\gamma_0-\gamma_3)\gamma_0 + i{\bbox{\alpha}}\cdot
  {\partial\over \partial {\bf y}}
  - {\bbox{\alpha}}\cdot\left( {\bf p} - p\, {\bf n}\right)\, .
  \label{a.6}
\end{eqnarray}
In the same coordinate system, $0.5\, (1-\gamma_3\gamma_0)\, 
v({\bf p}) = v({\bf p})$. Acting with (\ref{a.6}) on $v({\bf p})$,
we thus find to leading order in $E$ an expression proportional
to $\hat D\, v({\bf p})$. With these steps, the 
amplitude (\ref{a.2}) takes the form
%
%
\begin{eqnarray}
  I^{(N)}({\bf y}) &=&
  e^{-\, i\, p\, y_L}\, {\hat D}\, 
  {\cal P} \int \left( \prod_{i=1}^N\, d^3{\bf x}_i\,
    \Theta(x_i^L - x_{i-1}^L)\right)
  \nonumber \\
  && \times \left(\prod_{i=1}^N\, \lbrack -i\, A_0({\bf x}_i)\, 
    \rbrack\, 
    G_0({\bf x}_{i-1},{\bf x}_i|p)\right)\, \int d{\bf x}_\perp\,
  \nonumber \\
  && \times  
   G_0({\bf x}_N,{\bf x}|p)
   F_\infty({\bf x}_\perp,{\bf x}^L, {\bf p})\,
  v^{(r)}(p)\, .
  \label{a.7}
\end{eqnarray}
Here, we have introduced two elements of the abelian Furry approximation:
the outgoing transverse plane wave $F_\infty$ and the free 
Green's function $G_0$ of (\ref{2.15}) which we have 
identified here with the Gaussian ${\bf p}_i^\perp$-integrals,
\begin{eqnarray}
  &&G_0({\bf x}_{i-1}^\perp,x_{i-1}^L;{\bf x}_i^\perp,
                      x_i^L\vert p)
    \nonumber \\
  &&\qquad = \int {d^2{\bf p}_i^\perp\over (2\pi)^2}\, 
                    e^{i\, {\bf p}_i^{\perp}\cdot 
                      ({\bf x}_i^\perp - {\bf x}_{i-1}^\perp)}\,
                    e^{- i{{p_i^{\perp}}^2\over 2\, p}
                       (x_i^L - x_{i-1}^L)}\, .
  \label{a.8}
\end{eqnarray}
For the path-ordered product in (\ref{a.7}) which involves these 
Green's functions, we introduce the shorthand
\begin{eqnarray}
  &&\bar G^{(N)}({\bf y}_\perp,y_L;{\bf x}_\perp,x_L\vert p) =
  \nonumber \\
  && \quad {\cal P}\, \left( \prod_{i=1}^{N+1}\int d^3{\bf x}_i\, 
            \Theta(x_{i}^L - x_{i-1}^L)\right)\,
      G_0({\bf y}_\perp,y_L;{\bf x}_1^\perp,x_1^L\vert p)
  \nonumber \\
  && \times \left( \prod_{i=1}^{N}
     \lbrack -i\, A_0({\bf x}_i)\rbrack\, 
            G_0({\bf x}_i^\perp,x_i^L;{\bf x}_{i+1}^\perp,
                      x_{i+1}^L\vert p)\, \right)\, ,
  \label{a.9}
\end{eqnarray}
where ${\bf x}_{N+1}={\bf x}_\infty$. Allowing for arbitrary
many gluon exchanges, we have to sum over $N$.
\begin{equation}
  \bar G({\bf y}_\perp,y_L;{\bf x}_\perp,x_L\vert p) 
  =  \sum_{N=0}^\infty 
  \bar G^{(N)}({\bf y}_\perp,y_L;{\bf x}_\perp,x_L\vert p)\, .
  \label{a.10}
\end{equation}
From this one sees easily that $\bar G^{(N)}$ corresponds
exactly to the $N$-th order $O(A_0^N)$ term in (\ref{2.14}). 
The sum (\ref{2.12}) over $N$-fold scattering
diagrams takes the form of a non-abelian extension of
the Furry approximation.

We emphasize that the approximations used in our derivation and 
especially the different treatment of the quark propagators at 
the production vertex and in the final state rescattering part 
of the amplitude $I^{(N)}$ leads to a consistent high energy 
expansion with a norm accurate to leading order in $O(1/E)$ and a 
phase factor accurate up to order $O(1/E^2)$. The same
approximations were employed in recent calculations of rescattering 
amplitudes for which the production vertex $P$ is a photon or gluon 
emission vertex~\cite{GW94,BDMPS1}.

\section{$q\bar{q}$ Fock states}
\label{appb}

In this appendix, we give details of the derivation of 
transverse and longitudinal components of the squared incoming
Fock state
\begin{equation}
  \Phi(\Delta {\bf b}; \Delta {\bf \bar b};\alpha)
  = \Phi_\perp(\Delta {\bf b}; \Delta {\bf \bar b};\alpha)
    + \Phi_L(\Delta {\bf b}; \Delta {\bf \bar b};\alpha)\, .
    \label{b.1}
\end{equation}
We start with the longitudinal polarization $\epsilon_\mu^L$, 
satisfying $\epsilon^L\cdot k = 0$ and ${\epsilon^L}^2=1$,
\begin{equation}
  \epsilon_\mu^L = \frac{1}{Q}
      \left( \sqrt{\nu^2+Q^2}, \bbox{0}_\perp,\nu\right)\, .
      \label{b.2}
\end{equation}
The corresponding vertex function reads to leading order in
energy
\begin{equation}
  \Gamma^L = \epsilon_\mu^L\, \hat\Gamma^\mu
  = Q\, \sqrt{(1-\alpha)\, \alpha}\, r\, \delta_{r,r'}\, ,
    \label{b.3}
\end{equation}
where $r$, $r'$ are the helicities of the quark and antiquark.
The square of this longitudinal vertex function,
summed over the spin of the final state particles, reads
\begin{equation}
  \sum_{r,r'}
  \Gamma^L\, {\Gamma^L}^* = 2\, Q^2\, 
       (1-\alpha)\, \alpha\, .
  \label{b.4}
\end{equation}
This is a kinematical prefactor which factorizes in the integrand
of (\ref{3.7}). For the first interaction in
the photodissociation amplitude in (\ref{3.7}), we take some
longitudinal position $z_a$. The ${\bf y}$-integration in (\ref{3.7})
can then be done analytically. We find
\begin{eqnarray}
  &&I_L({\bf b}_2 - {\bf b}_1|y_L,z_a) 
  \nonumber \\
  && \qquad = \int {\it d}^2{\bf y}_\perp\, 
  \bar G_0({\bf b}_2,z_a;{\bf y}\vert p_2)\,  
  \bar G_0({\bf y};{\bf b}_1,z_a\vert p_1)
  \nonumber \\
  && \qquad = \frac{\mu}{2\pi\, i\, (z_a-y_L)}\, 
  \exp\Big\lbrace \frac{i\mu\, 
                   \left({\bf b}_2 - {\bf b}_1\right)^2    
                        }{2\, (z_a-y_L)}
                   \Big\rbrace\, ,
  \label{b.5}
\end{eqnarray}
where $\mu = (1-\alpha)\alpha\, \nu$. The virtual photon has to
dissociate at longitudinal position $y_L < z_a$ in order to make 
an interaction at $z_a$ possible. This limits the range of the
$y_L$-integral:
%
%
\begin{eqnarray}
  I_L(\Delta {\bf b}|z_a) 
  &=& \int\limits_{-\infty}^{z_a}\, dy_L\,
      I_L({\bf b}_2 - {\bf b}_1|y_L,z_a)
  \nonumber \\
  &=& \frac{(1-\alpha)\,\alpha\, \nu}{2\pi\, i}\, e^{i\, q\, z_a}\, 
  2\, K_0(\bar{\epsilon}\, |\Delta {\bf b}|)\, ,
   \label{b.6} \\
  \bar{\epsilon} &=& \sqrt{ (1-\alpha)\alpha\, Q^2 + m^2}\, .
   \label{b.7} 
\end{eqnarray}
If the first interaction occurs at $z_a$ in both the amplitude
and complex conjugate amplitude, then the square of the incoming
longitudinal $q\bar{q}$ Fock wavefunction is defined as the
combination of the squared emission vertex (\ref{b.4}) and the free
time evolution (\ref{b.5}), 
\begin{eqnarray}
  &&\Phi_L(\Delta {\bf b}; \Delta {\bf \bar b};\alpha)
  = \frac{\sum_{r,r'}\Gamma^L\, {\Gamma^L}^*}{4\nu^2\alpha(1-\alpha)}
    I_L(\Delta {\bf b}|z_a)\, I_L^*(\Delta {\bf \bar b}|z_a)
  \nonumber \\
  && \qquad = \frac{Q^2\, \alpha^2\, (1-\alpha)^2}{ 2\, (2\pi)^2}\, 
  4\, K_0(\bar{\epsilon}|\Delta {\bf b}|)\, 
  K_0(\bar{\epsilon}|\Delta {\bf \bar b}|)\, .
  \label{b.8}
\end{eqnarray}
This expression appears for longitudinal polarization in the
photodissociation cross section (\ref{3.9}).

For our calculations in section~\ref{sec4}, we have to consider
the more general case that the first interaction vertex occurs
at $z_a$ in the complex conjugated amplitude $M_{\rm fi}^\dagger$,
but at $z_b > z_a$ in the amplitude  $M_{\rm fi}$. This implies
the further free evolution of (\ref{b.5}) from $z_a$ to $z_b$,
\begin{eqnarray}
  &&I_L({\bf b}_2 - {\bf b}_1|y_L,z_b) 
  \nonumber \\
  && \qquad = \int {\it d}^2{{\bf b}_1'}\, 
  {\it d}^2{{\bf b}_2'}\, 
  G_0({\bf b}_2,z_b;{{\bf b}_2'},z_a\vert p_2)\,  
  \nonumber \\
  && \qquad \quad \times 
  I_L({{\bf b}_2'} - {{\bf b}_1'}|y_L,z_a)\,
  G_0({{\bf b}_1'},z_a;{\bf b}_1,z_b\vert p_1)
  \nonumber \\
  && \qquad = \int d{\bf r}_b(z_a)\, 
  K_0\left({\bf r}_b(z_b),z_b;{\bf r}_b(z_a),z_a|
      \nu\alpha(1-\alpha)\right)   
  \nonumber \\
  && \qquad \quad \times 
    I_L({\bf r}_b(z_a)|y_L,z_a)\, .
  \label{b.9}
\end{eqnarray}
Here, we have used the notation introduced in equations (\ref{4.16})
- (\ref{4.19}) with ${\bf r}_b(z_a) = {{\bf b}_2'} - {{\bf b}_1'}$ and
${\bf r}_b(z_b) = {\bf b}_2 - {\bf b}_1$. 
Equation (\ref{b.9}) allows us to circumvent a notational problem
which stems from the identity:
\begin{eqnarray}
  &&\int\limits_{-\infty}^{z_b}\, dy_L\,
  I(\Delta {\bf b}|y_L,z_b)
  \nonumber \\
  && \qquad =
  e^{iq\left(z_b-z_a\right)}\, 
  \int\limits_{-\infty}^{z_a}\, dy_L\,
  I(\Delta {\bf b}|y_L,z_a)\, .
  \label{b.10}
\end{eqnarray}
Equation (\ref{b.10}) shows that if the points of first 
interaction differ in $M_{\rm fi}^\dagger$ and $M_{\rm fi}$, then
phase factors $\exp\left(i\, q\, (z_b-z_a)\right)$ arise in
the squared free incoming wavefunction (\ref{b.8}). Hence, strictly 
speaking, we cannot do the $y_L$-integral {\it before} specifying 
the first points of interaction. Previous discussions of the
$q$-$\bar{q}$-dipole do not have this difficulty, since they
neglect these phase factors - a praxis which is justified 
for nuclear targets of size $L \ll 1/q$. If we want to keep
the phases $\exp\left(i\, q\, (z_b-z_a)\right)$ and yet
write the photoabsorption cross section (\ref{3.9}) in terms
of $\Phi_L$, we can use (\ref{b.9}) for a simple convention:
whenever the free Green's function 
$K_0\left({\bf r}_b(z_b),z_b;{\bf r}_b(z_a),z_a|     
\nu\alpha(1-\alpha)\right)$ acts on the first argument 
$\Delta {\bf b}$ of
$\Phi_L(\Delta {\bf b}; \Delta {\bf \bar b};\alpha)$,
the result is a phase  $\exp\left(i\, q\, (z_b-z_a)\right)$,
whenever it acts on the second argument, the result is the
complex conjugated phase. This allows us to expand the
cross section (\ref{4.23}) in powers of the opacity
without neglecting phase factors and without giving up
the simple representation of $\sigma^{\gamma^*\to q\bar{q}}$
in terms of $\Phi$.

We now turn to the transverse polarizations
\begin{eqnarray}
  \epsilon_\mu^\perp(\lambda) &=& \frac{1}{\sqrt{2}}
      \left(0,1,i\lambda,0\right)\, .
      \label{b.11}\\
  \Gamma^T_\lambda({\bf y}) &=& 
  \epsilon_\mu^\perp(\lambda)\Gamma^\mu 
  \nonumber \\
  &=&
  \frac{-1}{2\sqrt{(1-\alpha)\alpha}}
  \lbrack \delta_{r,r'}i\nabla_{\bf y}\cdot\epsilon^\perp(\lambda)
       \left( r'(1-2\alpha)+\lambda\right) 
  \nonumber \\
  && +
       \frac{m}{\sqrt{2}}\delta_{-r,r'} \left(1+\lambda r'\right)
       \rbrack\, .
       \label{b.12}
\end{eqnarray}
In contrast to the longitudinal case, this vertex does not factorize 
in the integrand of (\ref{3.7}). However, the ${\bf y}$-integration
can be done in analogy to (\ref{b.6}),
\begin{eqnarray}
  &&I^\perp_\lambda(\Delta {\bf b}|z_a) 
  \nonumber \\
  && \qquad = \int\limits_{-\infty}^{z_a} dy_L\, 
  \int {\it d}^2{\bf y}_\perp\, e^{i\, q\, y_L - \epsilon\, |y_L|}\, 
  G_0({\bf b}_2;{\bf y}\vert p_2)
  \nonumber \\
  && \qquad \qquad \times \Gamma^T_\lambda({\bf y}) 
  G_0({\bf y};{\bf b}_1\vert p_1)
  \nonumber \\
  && \qquad = \frac{(1-\alpha)\,\alpha\, \nu}{2\pi\, i}\, 
  e^{i\, q\, z_a}\, 
  2\, \Gamma^T_\lambda(\Delta {\bf b}) \, 
  K_0(\bar{\epsilon}\, |\Delta {\bf b}|)\, .
   \label{b.13}
\end{eqnarray}
The squared transverse wavefunction reads then 
\begin{eqnarray}
  && \Phi_\perp(\Delta {\bf b}; \Delta {\bf \bar b}; 
                 \alpha) 
      = \frac{\frac{1}{2} \sum_{\lambda,r,r'}}{4\nu^2\alpha(1-\alpha)}\,
        I^\perp_\lambda(\Delta {\bf b}|z_a)\, 
        {I^\perp_\lambda}^*(\Delta \bar {\bf b}|z_a)
   \nonumber \\
  && = \frac{\bar{\epsilon}^2}{2\, (2\pi)^2}\, 
  \frac{\Delta {\bf b}\cdot \Delta {\bf \bar b}}
        {|\Delta {\bf b}|\, |\Delta {\bf \bar b}|}
        K_1(\bar{\epsilon}{\Delta {\bf b}})\, 
        K_1(\bar{\epsilon}{\Delta \bar {\bf b}})\, 
        \left(\alpha^2+(1-\alpha)^2\right)
      \nonumber \\
  && \qquad + \frac{m^2}{2\, (2\pi)^2}\, 
     K_0(\bar{\epsilon}{\Delta {\bf b}})\, 
        K_0(\bar{\epsilon}{\Delta \bar {\bf b}})\, .
  \label{b.14}
\end{eqnarray}
For the case that the points of first interaction $z_a$ and
$z_b$ are different in the amplitude and complex conjugated
amplitude, the convention explained below (\ref{b.10}) 
applies. 

The arguments of the modified Bessel
functions $K_0$ and $K_1$ in (\ref{b.8}) and (\ref{b.14})
specify the transverse separation of  
the $q$-$\bar{q}$ dipole pair. This separation is given approximately
by $1/ \bar{\epsilon} =  1/\sqrt{ (1-\alpha)\alpha\, Q^2 + m^2}$. 
The higher the virtuality of the photon, the smaller the transverse
size of this dipole. In general, the approximation of the virtual 
photon by the $q$-$\bar{q}$ Fock state is reasonable at sufficiently 
high virtuality, $Q^2 \gg 1$ GeV$^2$, above the vector meson dominance
region~\cite{PW99}. There, the $q$-$\bar{q}$ separation is small enough 
to render the effects from string tension negligible and the quarks
undergo independent scatterings. For a recent attempt to model 
interactions between the quarks, see Ref.~\cite{KST99}.

\section{Classification of diagrams for (\ref{4.30})}
\label{appc}

\begin{figure}[h]\epsfxsize=5.7cm 
\centerline{\epsfbox{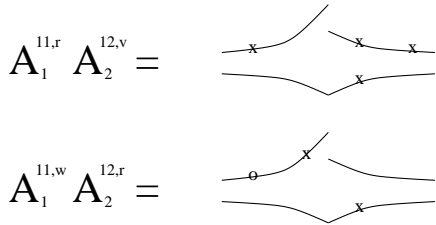}}
\vspace{0.5cm}
\caption{Example of our notation (\protect\ref{c.1}) for
diagrammatic contributions to the photoabsorption cross section.
For more details, see the text following (\protect\ref{c.1}). 
}\label{figc1}
\end{figure}

In this appendix, we use the identities Fig.~\ref{fig7}a and b to
simplify the diagrammatic contributions to the inelastic part of
the differential 
photoabsorption cross section (\ref{4.30}). For each $N$-th order term 
in the opacity expansion of (\ref{4.30}), we use the notational 
shorthand
\begin{equation}
  \prod_{i=1}^N\, A_i^{n_i,m_i,o}\, .
  \label{c.1}
\end{equation}
Here, the index $i$ labels the scattering potentials linked
to the $N$-th order term, $i=1$ being the first scattering 
potential. The superscripts $n_i,m_i,o$ denote how the $i$-th
scattering center is connected to the $q$-$\bar{q}$-system.
Each scattering center is linked twice to the $q$-$\bar{q}$-system,
once a momentum $\bbox{\kappa}_\perp$ flows into the system, once
it flows out. $n_i\in [1,2]$ specifies whether the inflowing
momentum is transfered to the upper quark line of momentum
$p_1$ ($n_i = 1$) or the lower quark line of momentum $p_2$
($n_i = 2$). Analogously, $m_i\in [1,2]$ specifies whether
the outflowing momentum comes from the upper ($m_i = 1$) or
lower ($m_i = 2$) quark line. The remaining superscript
$o \in [r,v,w]$ specifies the position of the two vertices from
the $i$-th interaction w.r.t. the cut: either both vertices
stand to the right of the cut ($o=v$) or
to the left of the cut ($o=w$), or they are a real contribution
($o=r$) where one vertex stands to the right and the other to
the left of the cut. We illustrate this notation with the
examples in Fig.~\ref{figc1}.

%
All $N$-th order contributions to the differential photoabsorption
cross section (\ref{4.30}) fall into exactly one of the following
three classes:
\begin{enumerate}
\item
There is at least one factor $A_j^{11,r}$ in the diagram (\ref{c.1}).
\item
There is at least one factor $A_j^{12,r}$ or $A_j^{21,r}$ in
the diagram (\ref{c.1}), but no factor $A_k^{11,r}$ (arbritrary $k$)
and no factor $A_k^{22,r}$ with $k < j$.
\item
(a) There is no factor $A_k^{11,r}$ (arbritrary $k$) but at least one 
factor $A_j^{12,r}$ or $A_j^{21,r}$ in the diagram (\ref{c.1}), and 
there is at least one factor $A_k^{22,r}$ with $k < j$.\\
or \\
(b)
There are no factors $A_j^{11,r}$, $A_j^{12,r}$ or $A_j^{21,r}$
in the diagram (\ref{c.1}).
\end{enumerate}
We now use the identities of Fig.~\ref{fig7}a,b to show that many
of the diagrams in these three classes cancel each other. In this
way we determine the only remaining contribution for each of the
three classes:

\underline{\it Class 1:} The only non-vanishing contributions 
are those which contain no terms 
$A_l^{22,o}$, $A_l^{12,o}$ or $A_l^{21,o}$ with $o\in [r,v,w]$.
They are shown in Fig.~\ref{fig10}(a1).
\\
{\it Argument:} Consider the subclass of diagrams containing 
factors $A_l^{22,o}$, $A_l^{12,o}$ or $A_l^{21,o}$. In case that
there is more than one such factor in the diagram, consider
the term with the largest index $l$, $l=k$ say. Leaving all
terms $j\not=k$ unchanged, we find in this class of diagrams
exactly one diagram with $A_k^{22,r}$,  $A_k^{22,v}$ and
$A_k^{22,w}$ on the $k$-th position. These three diagrams
cancel each other due to the identity Fig.~\ref{fig7}a.
Also, for all terms $j\not=k$ unchanged, we find exactly one
diagram with $A_k^{21,r}$ and $A_k^{21,v}$, which cancel due
to the identity Fig.~\ref{fig7}b. For the same reason, the
two diagrams with $A_k^{12,r}$ and $A_k^{12,w}$ cancel each
other (note that $A_k^{12,w}=A_k^{21,w}$ specifies the same
$k$-th term in the same diagram). 
As a consequence, no contribution which contains terms 
$A_l^{22,o}$, $A_l^{12,o}$ or $A_l^{21,o}$ gives a non-vanishing
contribution.

\underline{\it Class 2:} The only non-vanishing contributions 
contain exactly one real term $A_j^{12,r}$ or $A_j^{21,r}$
with no real term $A_k^{11,r}$ or $A_k^{22,r}$, $k$ arbitrary,
and with no contact terms $A_k^{12,v}$, $A_k^{12,w}$, 
$A_k^{22,v}$, or $A_k^{22,w}$ for $k>j$.
They are shown in Fig.~\ref{fig10}(a2).\\
{\it Argument:} Choose in each diagram the term $A_j^{12,r}$ 
or $A_j^{21,r}$ with the lowest index $j$. Consider 
diagrams which contain terms $A_k^{22,o}$, $A_k^{12,o}$
or $A_k^{21,o}$, with $k>j$ and $o=r,v,w$. Taking $k$ maximal 
and leaving all other terms unchanged, these diagrams cancel 
due to the identities Fig.~\ref{fig7}a and b. 
   
\underline{\it Class 3:} The only non-vanishing contributions 
contain exactly one real term $A_j^{22,r}$ but no real terms
$A_j^{11,r}$, $A_j^{12,r}$, or $A_j^{21,r}$ and no
contact terms $A_k^{22,v}$, $A_k^{22,w}$, $A_k^{12,v}$ or
$A_k^{12,w}$ with $k > j$. 
They are shown in Fig.~\ref{fig10}(a3).\\
{\it Argument:} Consider in each diagram the term $A_j^{22,r}$
with lowest index $j$. Look then at the largest index $k>j$
linking to the $p_2$-line. Leaving all other terms unchanged,
the sum of the contributions with different configuration at 
position $k$ ensures
cancellation: $A_k^{22,r}$, $A_k^{22,v}$ and $A_k^{22,w}$
cancel each other due to identity Fig.~\ref{fig7}a.
$A_k^{21,r}$, $A_k^{21,v}$ and $A_k^{12,r}$, $A_k^{21,w}$
cancel each other due to identity Fig.~\ref{fig7}b. Thus 
only contributions with exactly one real term $A_j^{22,r}$
survive cancellation.


\end{document}